\documentclass[aps,prfluids,preprint,onecolumn,preprintnumbers,superscriptaddress,nofootinbib,longbibliography]{revtex4-2}

\usepackage{amsmath,amssymb,bm}
\usepackage{mathrsfs}
\usepackage{graphicx}
\usepackage{booktabs}
\usepackage{microtype}
\usepackage{etoolbox}
\usepackage{pgfplots}
\pgfplotsset{compat=1.18}
\pgfplotsset{paper axis/.style={
  scale only axis, axis on top,
  axis background/.style={fill=white},
  axis line style={line width=0.6pt},
  tick style={black!55,line width=0.6pt},
  major tick length=3pt, minor tick length=1.8pt,
  tick align=outside, xtick pos=bottom, ytick pos=left,
  grid=none, font=\small,
  label style={font=\normalsize},
  tick label style={font=\small},
  every axis plot/.append style={line width=1.5pt},
  legend style={font=\small,draw=none,fill=none,cells={anchor=west},column sep=3pt},
  clip=true, clip mode=individual,
  scaled ticks=false,
}}
\definecolor{paperBlue}{HTML}{1D6996}
\definecolor{paperGreen}{HTML}{0F8554}
\definecolor{paperRed}{HTML}{CC503E}

\definecolor{dblue}{HTML}{2e3092}
\usepackage{hyperref}
\hypersetup{colorlinks=true,linkcolor=dblue,urlcolor=dblue,citecolor=dblue,filecolor=dblue}

\makeatletter
\patchcmd{\frontmatter@abstract@produce}
  {\preprintsty@sw}{\@secondoftwo}
  {}{\PackageError{manuscript}{REVTeX abstract pagination patch failed}{Check the REVTeX version.}}
\makeatother

\newcommand{\dd}{\mathrm{d}}
\newcommand{\Dt}{\mathrm{D}_t}
\newcommand{\eps}{\varepsilon}
\newcommand{\bfx}{\bm{x}}
\newcommand{\bfv}{\bm{v}}
\newcommand{\bfu}{\bm{u}}
\newcommand{\bfg}{\bm{g}}
\newcommand{\bfa}{\bm{a}}

\begin{document}
\title{Preferential concentration of cloud droplets: Coupled effects of phase change and gravitational settling}
\author{Zi-Mo Liao}
\email{zimoliao@mail.ustc.edu.cn}
\affiliation{Department of Modern Mechanics, University of Science and Technology of China, Hefei 230027, China}
\date{October 13, 2026}

\begin{abstract}
Cloud droplets respond to turbulent motion with finite inertia and settle under gravity, while condensation and evaporation change their sizes.
The nonuniform accumulation of cloud droplets in turbulent flows, known as preferential concentration, has attracted sustained interest because it affects droplet collision and coalescence rates.
We develop an analytical theory that quantifies the coupled effects of droplet inertia, gravitational settling, and phase change on concentration changes.
Applying this theory to simulations of two-dimensional homogeneous thermal convection reveals that phase change markedly enhances preferential concentration in this flow, primarily through spatial differences in settling speed.
These differences arise because droplets experience different thermodynamic histories along their trajectories and thus acquire different sizes.
Droplet transport further increases this settling contribution by rearranging droplets of different sizes.
Fluid deformation partly offsets this enhancement by thinning concentration structures, reducing the preferential concentration measured at a fixed spatial scale.
\end{abstract}
\maketitle

\section{Introduction}
Inertial particles, such as cloud droplets, accumulate nonuniformly in turbulent flows, a phenomenon known as preferential concentration \cite{Bec2024}.
Cloud droplets also settle under gravity and undergo phase change---condensation and evaporation---which continually changes their sizes.
These processes are coupled: changes in droplet size alter both gravitational settling and the inertial response to the turbulent flow, thereby modifying preferential concentration \cite{Pinsky1999,Vaillancourt2002,Sardina2015}.
This coupled evolution of droplet sizes and their spatial distribution affects collision and coalescence rates \cite{Devenish2012,GrabowskiWang2013}.
Its effects extend to warm-rain formation \cite{GrabowskiWang2013} and cloud entrainment \cite{Mellado2017}, illustrating how small-scale processes influence large-scale cloud dynamics \cite{Bodenschatz2010}.

For small heavy particles of fixed size, Maxey's small-inertia analysis shows that particles accumulate in strain-dominated regions \cite{Maxey1987}, providing a theoretical basis for preferential concentration \cite{Chun2005,Bec2024}.
\citet{Bec2014} and \citet{Ireland2016Gravity} showed that gravitational settling can enhance or weaken preferential concentration of monodisperse particles, depending on particle inertia and the relative strength of gravity.
For particles of different sizes, \citet{Dhanasekaran2022} showed that differences in settling speed can weaken preferential concentration.
For phase-changing droplets, sizes not only differ among droplets but also evolve over time \cite{Sardina2015,Li2021}.
\citet{Sahu2018} observed that evaporation can enhance preferential concentration in a turbulent spray.
\citet{Elperin2013,Elperin2015} developed a theory of droplet clustering in temperature-stratified turbulence.
\citet{Avni2023Mass,Avni2023Heat} showed that condensation and evaporation can cause droplets to accumulate at the edge of a region within a Burgers vortex where droplets grow by condensation.
Using a small-inertia expansion, \citet{Dagan2017} quantified droplet spacing in an oscillatory vortex without gravity, assuming that the squared droplet diameter decreases at a constant rate during evaporation.
However, an analytical description of how droplet inertia, gravitational settling, and phase change jointly determine preferential concentration in general cloud turbulence, with their contributions made explicit, is still lacking.

Phase-changing droplets acquire different sizes as they experience different thermodynamic histories---i.e., histories of temperature and water-vapor mixing ratio---along their trajectories \cite{Celani2005,Sardina2015,Siewert2017,Fries2023}.
How these size differences affect preferential concentration depends on how droplets of different sizes are distributed in space.
For example, differences in settling speed bring droplets closer together when faster-settling droplets lie above slower-settling droplets, whereas the opposite ordering drives them apart.
Advection and settling continually change these relative positions.
The key is therefore to connect thermodynamic histories and droplet transport to the resulting spatial differences in droplet size and, in turn, to concentration changes.
To this end, we extend Maxey's small-inertia analysis to phase-changing droplets and derive a concentration law.
We quantify preferential concentration using the relative entropy of the spatially averaged concentration field and use the concentration law to separate the physical contributions to its evolution.
In simulations of two-dimensional homogeneous thermal convection, we find that phase change substantially enhances preferential concentration, primarily through the spatial differences in settling speed it generates.
We further identify how droplet transport modifies the settling contribution and how fluid deformation affects the relative entropy.

\section{Concentration law for cloud droplets}
Let $\bfu(\bfx,t)$ be the incompressible fluid velocity, $\Dt=\partial_t+\bfu\cdot\nabla$ its material derivative, and $\bfa=\Dt\bfu$ its acceleration.
For a dilute suspension of spherical water droplets with diameters much smaller than the Kolmogorov length scale and mass density much larger than that of air \cite{GrabowskiWang2013}, the droplet position $\bfx_p$, velocity $\bfv_p$, and internal state $\bm{\xi}_p$ obey \cite{Bec2024}
\begin{equation}
 \dot{\bfx}_p=\bfv_p,\qquad
 \dot{\bfv}_p=\frac{\bfu(\bfx_p,t)-\bfv_p}{\tau_p}+\bfg,\qquad
 \dot{\bm{\xi}}_p=\bm{\mathcal R}(\bm{\xi}_p,\bfx_p,\bfv_p,t).
 \label{eq:particle}
\end{equation}
Here $\bm{\xi}_p$ includes squared radius $r^2$ and, if evolved independently, droplet temperature; $\bm{\mathcal R}$ specifies their rates of change.
For water density $\rho_w$, air density $\rho_a$, and air kinematic viscosity $\nu$,
the droplet response time is $\tau_p=2\rho_wr^2/(9\rho_a\nu)$ and $\bfg=(1-\rho_a/\rho_w)\bfg_0$, where $\bfg_0$ is gravitational acceleration.
Following \citet{MacMillan2022}, we model droplet motion using Stokes drag and gravity, neglecting droplet--droplet collisions.
The following derivation considers intervals without droplet creation or disappearance.

To obtain a small-inertia expansion, we embed the droplet equations in a one-parameter family by replacing $\tau_p$ with $\tau_p^{(\eps)}=\eps\tau_p$ \cite{BenderOrszag1999}.
We let $\eps\to0$ while holding the flow and phase-change model fixed; $\eps=1$ recovers the original equations.
The scaled response time $\tau_p^{(\eps)}$ is evaluated using the evolving droplet radius.
We seek an approximation on a finite time interval, fixed as $\eps\to0$, after the initial velocity transient has decayed.
Droplet radii remain positive, and response times are short compared with the timescales of flow variation and droplet-size evolution.
After this transient, the droplet acceleration remains bounded, so the momentum equation gives $\bfv_p-\bfu=O(\eps)$.
The leading-order balance then gives \cite{Maxey1987,Haller2008}
\begin{equation}
 \bfv_p(t)=\bfu(\bfx_p(t),t)
 +\tau_p^{(\eps)}(t)\left[\bfg-\bfa(\bfx_p(t),t)\right]
 +O(\eps^2).
\label{eq:slow}
\end{equation}
The error estimate and its assumptions are given in the Supplemental Material \cite{SupplementalMaterial}.

Droplets at the same position may have different velocities and internal states.
For a continuum droplet population, let $f(\bfx,\bfv,\bm\xi,t)$ denote its phase-space distribution, transported by the particle equations in a given realization of the flow \cite{Fox2012,Liao2025}.
Here $\bfx$, $\bfv$, and $\bm\xi$ are the position, velocity, and internal-state coordinates, respectively.
We assume smooth spatial moments and define the number density $c$, local mean velocity $\langle\bfv\rangle$, and local mean response time $\langle\tau\rangle$ by
\begin{align}
 c=\int f\,\dd\bfv\,\dd\bm\xi,\qquad
 c\langle\bfv\rangle=\int\bfv f\,\dd\bfv\,\dd\bm\xi,\qquad
 c\langle\tau\rangle=\int\tau_p^{(\eps)}f\,\dd\bfv\,\dd\bm\xi.
 \label{eq:meanresponse}
\end{align}
Angle brackets denote averages over velocity and internal state, weighted by droplet number at each position and time.
Number conservation and the average of Eq.~\eqref{eq:slow} give
\begin{align}
 \partial_t c+\nabla\cdot(c\langle\bfv\rangle)&=0,\label{eq:numberconservation}\\
 \langle\bfv\rangle&=\bfu+\langle\tau\rangle(\bfg-\bfa)+O(\eps^2).
 \label{eq:momentflux}
\end{align}
Define the fluid velocity gradient tensor $\mathsf{A}$ with components $A_{ij}=\partial_j u_i$, the strain-rate tensor $\mathsf{S}=(\mathsf{A}+\mathsf{A}^T)/2$, and the rotation-rate tensor $\mathsf{\Omega}=(\mathsf{A}-\mathsf{A}^T)/2$.
For incompressible flow, the second invariant of $\mathsf{A}$ is $Q=(\Omega_{ij}\Omega_{ij}-S_{ij}S_{ij})/2$ \cite{Chong1990}.
We use the Einstein summation convention for repeated indices.
Incompressibility gives $\nabla\cdot\bfa=\operatorname{tr}(\mathsf{A}^2)=-2Q$.
Substituting Eq.~\eqref{eq:momentflux} into Eq.~\eqref{eq:numberconservation} and expanding the divergence therefore gives
\begin{equation}
 (\partial_t+\langle\bfv\rangle\cdot\nabla)\ln c
 =-\nabla\cdot\langle\bfv\rangle
 =\underbrace{-2\langle\tau\rangle Q}_{\mathcal C_Q}
  \underbrace{-\bfg\cdot\nabla\langle\tau\rangle}_{\mathcal C_g}
  \underbrace{+\bfa\cdot\nabla\langle\tau\rangle}_{\mathcal C_a}
  +O(\eps^2).
 \label{eq:concentration}
\end{equation}

For droplets with the same constant response time, Eq.~\eqref{eq:concentration} reduces to Maxey's result for preferential concentration in strain-dominated regions, where $Q<0$ \cite{Maxey1987}.
In this case, gravitational settling is spatially uniform and produces no local convergence or divergence \cite{Shaw2003,Gustavsson2014}.
When the local mean response time varies in space, gravity and fluid acceleration contribute through $\mathcal C_g$ and $\mathcal C_a$, respectively.
For $\bfg=-g\bm e_y$, where $g>0$ and $\bm e_y$ is the upward unit vector, the settling contribution is $g\partial_y\langle\tau\rangle$: it increases concentration when droplets above settle faster on average than those below, even when $Q=0$, and decreases it for the opposite ordering.

Using Kolmogorov scales, we estimate the settling-to-$Q$ ratio as
\begin{equation}
 \frac{|\mathcal C_g|}{|\mathcal C_Q|}
 =\frac{g|\partial_y\ln\langle r^2\rangle|}{2|Q|}
 \sim\frac{g\nu}{\epsilon_\nu}|\partial_y\ln\langle r^2\rangle|,
 \label{eq:settlingQscale}
\end{equation}
where $\epsilon_\nu$ is the kinetic-energy dissipation rate.
Table~\ref{tab:cloudscales} gives the size differences required for comparable contributions in warm-cloud conditions \cite{GrabowskiWang2013}, using $\nu=1.57\times10^{-5}\,\mathrm{m^2\,s^{-1}}$ and $g=9.81\,\mathrm{m\,s^{-2}}$.
At $\epsilon_\nu=10^{-4}\,\mathrm{m^2\,s^{-3}}$, a $1\%$ increase in $\langle r^2\rangle$ over $1\,\mathrm{cm}$ already gives a gravity-to-$Q$ ratio of about $1.5$.
At $10^{-2}\,\mathrm{m^2\,s^{-3}}$, the same gradient gives only $0.015$.
For fluid acceleration,
\begin{equation}
 \frac{|\mathcal C_a|}{|\mathcal C_Q|}
 =\frac{|\bm a\cdot\nabla\ln\langle r^2\rangle|}{2|Q|}
 \sim\eta|\nabla\ln\langle r^2\rangle|,
 \label{eq:accelerationQscale}
\end{equation}
where $\eta=(\nu^3/\epsilon_\nu)^{1/4}$ is the Kolmogorov length scale.
The actual ratio also depends on the alignment of acceleration with the size gradient.
For the same $1\%$ variation over $1\,\mathrm{cm}$, the characteristic acceleration-to-$Q$ ratios are $0.0025$ and $0.00079$ at the two dissipation rates above.
Thus, in weak cloud turbulence, a modest vertical size gradient can make settling as important as the classical strain--rotation mechanism for local concentration changes, while fluid acceleration contributes much less.
Notably, recent cloud observations and simulations have identified regions with different droplet size distributions \cite{Allwayin2024,Allwayin2025,Yang2025} and revealed pronounced preferential concentration \cite{Thiede2026}.
Our concentration law, Eq.~\eqref{eq:concentration}, quantitatively connects spatial differences in droplet size to concentration changes, with explicit contributions from local strain and rotation, gravitational settling, and fluid acceleration.

\begin{table}[!t]
\caption{Estimated size gradients for $|\mathcal C_g|\sim|\mathcal C_Q|$ in warm clouds. In the last column, $\Delta\langle r^2\rangle$ is the difference in local mean squared radius between two points separated vertically by $1\,\mathrm{cm}$, and $\langle r^2\rangle_{\min}$ is the smaller of the two values.}
\label{tab:cloudscales}
\begin{ruledtabular}
\begin{tabular}{lccc}
Cloud conditions & $\epsilon_\nu\ (\mathrm{m^2\,s^{-3}})$ & $|\partial_y\ln\langle r^2\rangle|\ (\mathrm{m^{-1}})$ & $\Delta\langle r^2\rangle/\langle r^2\rangle_{\min}$\\
\midrule
Stratocumulus \cite{Siebert2010} & $10^{-4}$ & $0.65$ & $0.65\%$\\
Stratocumulus \cite{GrabowskiWang2013} & $10^{-3}$ & $6.5$ & $6.5\%$\\
Cumulus \cite{GrabowskiWang2013} & $10^{-2}$ & $65$ & $65\%$
\end{tabular}
\end{ruledtabular}
\end{table}

\section{Droplet response time from thermodynamic history}

The concentration law depends on both the local mean response time and its spatial gradient.
We now determine how these quantities develop from the thermodynamic histories and motion of the droplets.
We follow the droplet heat- and mass-transfer model of \citet{MacMillan2022}, specialized to pure water, and adopt the quasi-steady approximation for droplet temperature \cite{Vaillancourt2001}.
Condensation and evaporation are driven by the difference between the ambient water-vapor mixing ratio $q$ and its equilibrium value at the droplet surface $q_*$.
For a droplet of radius $r$ and temperature $T_p$ in air at temperature $T$, this equilibrium value is \cite{Richter2021}
\begin{equation}
 q_*(r,T_p,T)=\frac{M_we_s(T)}{R_uT_p\rho_a}
 \exp\left[\frac{L_vM_w}{R_u}\left(\frac1T-\frac1{T_p}\right)
 +\frac{2\sigma M_w}{R_u\rho_wrT_p}\right].
 \label{eq:surfaceequilibrium}
\end{equation}
Here $e_s(T)$ is the saturation vapor pressure over a plane water surface, evaluated using the modified Magnus formula \cite{Alduchov1996}.
The constants $M_w$, $R_u$, $L_v$, and $\sigma$ are the molar mass of water, universal gas constant, latent heat of vaporization, and surface tension, respectively.
The two terms in the exponent account for the temperature difference and surface curvature, respectively (see Sec.~II~A of the Supplemental Material for details).

For the physical system ($\eps=1$), the radius and response time evolve according to
\begin{equation}
 \frac{\mathrm d r^2}{\mathrm dt}=\frac{\rho_a D_v\mathrm{Sh}}{\rho_w}(q-q_*),\qquad
 \dot\tau_p=\frac{2\mathrm{Sh}}{9\mathrm{Sc}}(q-q_*).
 \label{eq:growth}
\end{equation}
Here $D_v$ is the water-vapor diffusivity and $\mathrm{Sc}=\nu/D_v$ is the Schmidt number.
The Sherwood number for mass transfer is $\mathrm{Sh}=2+0.6\mathrm{Re}_p^{1/2}\mathrm{Sc}^{1/3}$ \cite{Ranz1952I,Ranz1952II}, where $\mathrm{Re}_p=2r|\bfv_p-\bfu|/\nu$ is the droplet Reynolds number based on the slip velocity.
The droplet temperature $T_p$ entering $q_*$ is determined by balancing heat exchange with the air against latent heat released or absorbed by phase change \cite{Richter2021,Vaillancourt2001}:
\begin{equation}
 c_{pa}\frac{\mathrm{Nu}}{\mathrm{Pr}}(T-T_p)+L_v\frac{\mathrm{Sh}}{\mathrm{Sc}}\left[q-q_*(r,T_p,T)\right]=0.
 \label{eq:dropletheatbalance}
\end{equation}
Here $c_{pa}$ is the specific heat capacity of air at constant pressure.
The Nusselt number for heat transfer is $\mathrm{Nu}=2+0.6\mathrm{Re}_p^{1/2}\mathrm{Pr}^{1/3}$ \cite{Ranz1952I,Ranz1952II}, where $\mathrm{Pr}=\nu/\kappa_T$ is the Prandtl number and $\kappa_T$ is the thermal diffusivity of air.
Since $T_p$ is determined algebraically, the evolving internal state is $\bm\xi_p=r_p^2$.

These equations link the evolution of droplet size and response time to thermodynamic history.
In cloud turbulence, droplets encounter different temperature and water-vapor fluctuations along their trajectories and therefore grow or evaporate at different rates.
To obtain the evolution of the local mean response time, we multiply the phase-space transport equation by $\tau_p^{(\eps)}$ and integrate over velocity and internal state:
\begin{equation}
 \partial_t\bigl(c\langle\tau\rangle\bigr)
 +\nabla\cdot\bigl(c\langle\tau\bfv\rangle\bigr)
 =c\langle\dot\tau\rangle.
 \label{eq:responsemoment}
\end{equation}
Here $\langle\dot\tau\rangle=c^{-1}\int\dot\tau_p^{(\eps)}f\,\dd\bfv\,\dd\bm\xi$ is the local mean rate of response-time change due to phase change.
The flux $c\langle\tau\bfv\rangle=\int\tau_p^{(\eps)}\bfv f\,\dd\bfv\,\dd\bm\xi$ describes transport of response time by the droplets.
Expanding the time derivative in Eq.~\eqref{eq:responsemoment} and using number conservation, Eq.~\eqref{eq:numberconservation}, gives
\begin{equation}
 \partial_t\langle\tau\rangle
 =\langle\dot\tau\rangle
 \underbrace{-\frac{\nabla\cdot(c\langle\tau\bfv\rangle)
 -\langle\tau\rangle\nabla\cdot(c\langle\bfv\rangle)}{c}}_{\mathcal T_\tau}.
 \label{eq:meanresponseexact}
\end{equation}
The first term describes changes in the local mean response time due to condensation and evaporation.
The second term, $\mathcal T_\tau$, describes changes in the local mean response time caused by droplet transport (see Sec.~I~B of the Supplemental Material for detailed derivations).

\section{Preferential concentration in thermal convection}
We use two-dimensional homogeneous thermal convection \cite{Celani2002,Mazzino2017,Winchester2024} to examine how phase change modifies preferential concentration.
The flow is driven by a constant vertical temperature gradient in a doubly periodic domain.
The governing equations are
\begin{equation}
\begin{aligned}
 \partial_t\bfu+\bfu\cdot\nabla\bfu&=-\rho_a^{-1}\nabla p+\nu\nabla^2\bfu
       +\mu\nabla^{-2}\bfu+g_0\beta\theta\bm e_y,
       \qquad\nabla\cdot\bfu=0,\\
 \partial_t\theta+\bfu\cdot\nabla\theta&=\kappa_T\nabla^2\theta+\frac{\Delta T}{L}u_y.
\end{aligned}
\label{eq:hrbflow}
\end{equation}
These equations use the Oberbeck--Boussinesq approximation, in which density variations are retained only in the buoyancy term.
Here $p$ is the pressure, $\theta$ is the temperature fluctuation, and $\Delta T$ is the mean temperature difference across the domain height $L$.
The thermal expansion coefficient is $\beta=1/(300\,\mathrm K)$.
Gravity has magnitude $g_0=|\bfg_0|$.
Large-scale dissipation is provided by $\mu\nabla^{-2}\bfu$ \cite{Winchester2024}, where $\mu>0$ and $\nabla^{-2}$ acts only on nonzero Fourier modes.

The domain is a square of side $L=0.20\,\mathrm m$.
We adopt the air properties $\nu=1.57\times10^{-5}\,\mathrm{m^2\,s^{-1}}$, $\mathrm{Pr}=\nu/\kappa_T=0.71$, and $\mathrm{Sc}=0.615$ from Table II of \citet{MacMillan2022}.
We choose $\Delta T=4\,\mathrm K$ to produce appreciable changes in droplet size over the two-second observation interval.
The resulting Rayleigh number is $Ra=g_0\beta\Delta TL^3/(\nu\kappa_T)=3.01\times10^6$.
The free-fall velocity is $U_f=(g_0\beta\Delta TL)^{1/2}$ and the free-fall time is $t_f=L/U_f=1.237\,\mathrm s$.
Following \citet{Winchester2024}, we set the large-scale dissipation coefficient to $\mu=(2\pi)^{5/2}U_f/L^3$.
The water-vapor mixing ratio $q$ is held uniform at its saturation value over a plane water surface at $300\,\mathrm K$.
With $q$ fixed, temperature variations change the surface equilibrium value $q_*$ and hence the difference $q-q_*$ that drives condensation and evaporation.
We initialize the flow with a multimode velocity perturbation and allow thermal convection to develop for $20t_f$ before releasing particles.
At release, the Reynolds number is $\mathrm{Re}=u_{\mathrm{rms}}L/\nu\simeq545$, where $u_{\mathrm{rms}}=[L^{-2}\int_0^L\int_0^L|\bfu|^2\,\mathrm dx\,\mathrm dy]^{1/2}$.
The viscous dissipation rate at release is $\epsilon_\nu\simeq2.28\times10^{-4}\,\mathrm{m^2\,s^{-3}}$, within the range observed in weakly turbulent clouds \cite{GrabowskiWang2013}.

We release $N=512^2$ fixed-radius particles and the same number of phase-changing droplets at identical, uniformly spaced positions.
Time $t$ is measured from particle release.
All particles initially move with the local fluid velocity and have radius $r_0=15\,\mu\mathrm m$, corresponding to a response time of $2.654\,\mathrm{ms}$ and a Stokes settling speed of $0.0260\,\mathrm{m\,s^{-1}}$.
The initial Stokes number is $\mathrm{St}_{\eta,0}=\tau_p(0)/\tau_\eta(0)\simeq0.0101$, where $\tau_\eta=(\nu/\epsilon_\nu)^{1/2}$ is the viscous dissipation timescale.
We consider one-way coupling, in which the droplets do not affect the flow or thermodynamic fields.
Both populations obey the position and velocity equations in Eq.~\eqref{eq:particle}.
The fixed-radius particles retain $r=r_0$, whereas the radii of phase-changing droplets evolve according to Eqs.~\eqref{eq:surfaceequilibrium}--\eqref{eq:dropletheatbalance}.
The ambient temperature $T$ in Eqs.~\eqref{eq:surfaceequilibrium} and \eqref{eq:dropletheatbalance} is evaluated along each droplet trajectory as $T=T_p^{\rm amb}(t)$, where
\begin{equation}
 T_p^{\rm amb}(t)=300\,\mathrm K+\Delta T/2-\frac{\Delta T}{L}y_p^{\rm unwrapped}(t)
                    +\theta(\bfx_p(t),t).
 \label{eq:sampledtemperature}
\end{equation}
Here $y_p^{\rm unwrapped}(t)=y_p(0)+\int_0^t v_{p,y}(t')\,\mathrm dt'$ is the droplet's vertical coordinate accumulated continuously across periodic boundaries, with $y_p(0)\in[0,L)$.
We use this continuous vertical coordinate to evaluate the mean temperature and map particle positions into the periodic domain for flow interpolation and concentration statistics.
Comparisons with concentration statistics evaluated at the actual droplet heights confirm that periodic projection does not change the main qualitative conclusions below (see Sec.~II~C of the Supplemental Material).

We solve the flow equations using the pseudospectral method with $128\times128$ Fourier modes and $3/2$-rule dealiasing \cite{Peyret2002}.
Time integration uses a third-order Runge--Kutta scheme in integrating-factor form \cite{Lawson1967}.
Particle motion and droplet mass are advanced together using a backward-Euler scheme \cite{MacMillan2022}.
The flow and particle equations use the same time step of $0.5\,\mathrm{ms}$.
Fluid fields are interpolated to particle positions using four-point Lagrange interpolation in each coordinate direction.
Details of the numerical implementation are given in the Supplemental Material \cite{SupplementalMaterial}.

\begin{figure}[!hp]
 \centering
 \input{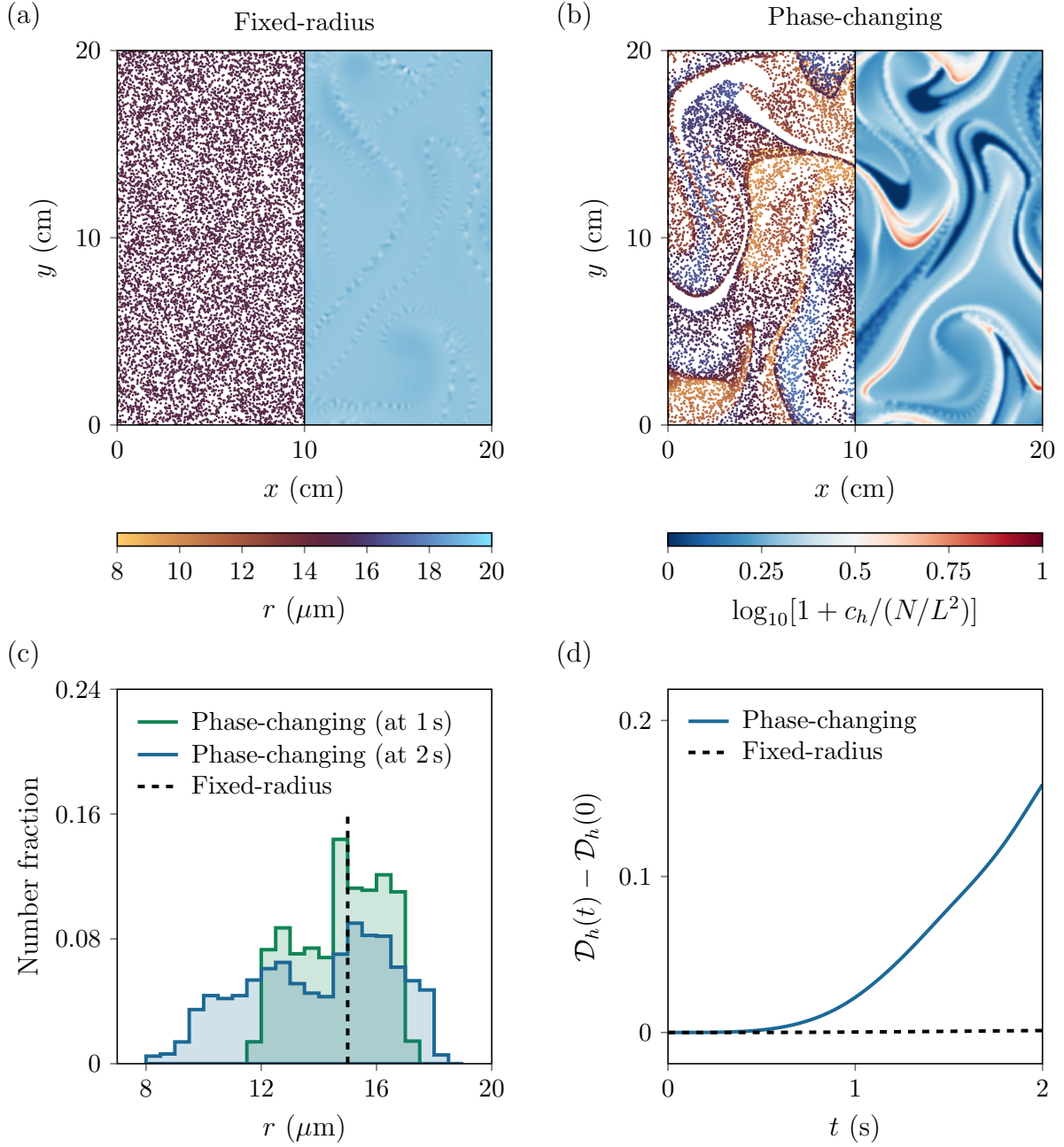}
 \setlength{\abovecaptionskip}{0pt}
 \caption{(a,b) Spatial distributions of fixed-radius particles and phase-changing droplets at $t=2\,\mathrm s$ after release. In each panel, the left half shows a random sample colored by radius ($20\,000$ particles sampled over the full domain), and the right half shows $\log_{10}[1+c_h/(N/L^2)]$, with concentration $c_h$ reconstructed from all particles using Eq.~\eqref{eq:kernelnumberdensity}. (c) Droplet size distributions at $1$ and $2\,\mathrm s$, shown as number fractions in $0.5\,\mu\mathrm m$ bins; the vertical dashed line marks the fixed radius of $15\,\mu\mathrm m$. (d) Relative-entropy increase for the two populations, $\mathcal D_h(t)-\mathcal D_h(0)$ [Eq.~\eqref{eq:concentrationentropy}].}
 \label{fig:hrbgeometry}
\end{figure}

\emph{Size dispersion and preferential concentration.}
Figures~\ref{fig:hrbgeometry}(a,b) show the spatial distributions of fixed-radius particles and phase-changing droplets at $t=2\,\mathrm s$.
The fixed-radius particles remain more uniformly distributed, whereas the phase-changing droplets form regions of high concentration separated by voids.
Figure~\ref{fig:hrbgeometry}(c) shows that condensation and evaporation broaden the initially monodisperse size distribution over this two-second interval, suggesting that the resulting size differences are responsible for the enhanced preferential concentration.

To quantify preferential concentration, we reconstruct the number density from particle positions using a normalized, nonnegative kernel $W_h$.
We construct $W_h$ by taking the product of one-dimensional quintic B-spline kernels \cite{Price2012} in the $x$ and $y$ directions, with the same kernel width $h$.
The reconstructed number density is
\begin{equation}
 c_h(\bfx,t)=\sum_{p=1}^N W_h(\bfx-\bfx_p(t)).
 \label{eq:kernelnumberdensity}
\end{equation}
We choose $h=1.276\,\mathrm{mm}$, large enough to average over many particles but small enough to distinguish regions of high concentration from voids.
The resulting concentration fields are shown in the right halves of Figs.~\ref{fig:hrbgeometry}(a,b).
We measure the departure from uniform concentration using the relative entropy
\begin{equation}
 \mathcal D_h(t)=\frac{1}{N}\int_0^L\!\int_0^L
 c_h(x,y,t)\ln\!\left[\frac{L^2c_h(x,y,t)}{N}\right]\,\dd x\,\dd y.
 \label{eq:concentrationentropy}
\end{equation}
Uniform concentration gives $\mathcal D_h=0$; larger values indicate stronger concentration nonuniformity at fixed $h$.
As shown in Fig.~\ref{fig:hrbgeometry}(d), the relative entropy increases by $0.1589$ for phase-changing droplets and $0.001246$ for fixed-radius particles by $t=2\,\mathrm s$.
This marked difference is consistent with the stronger preferential concentration of phase-changing droplets evident in Figs.~\ref{fig:hrbgeometry}(a,b).

\emph{Contributions to preferential concentration.}
To identify the physical processes responsible for this enhancement, we derive a relative-entropy budget from the concentration law, Eq.~\eqref{eq:concentration}.
For any droplet quantity $\psi_p$, the corresponding local mean is
$\langle\psi\rangle_h=c_h^{-1}\sum_p\psi_pW_h(\bfx-\bfx_p)$.
Number conservation gives $\partial_t c_h+\nabla\cdot(c_h\langle\bfv\rangle_h)=0$, and integration by parts over the periodic domain yields
\begin{equation}
 \frac{\mathrm d\mathcal D_h}{\mathrm dt}
 =-\frac{1}{N}\int_0^L\!\int_0^L c_h\nabla\cdot\langle\bfv\rangle_h\,\mathrm dx\,\mathrm dy.
 \label{eq:entropyconcentration}
\end{equation}
The concentration law separates local convergence into contributions from strain and rotation, gravitational settling, and fluid acceleration.
Integrating each contribution over space with weight $c_h/N$, and then over time, gives
\begin{equation}
 \mathcal D_\alpha(t)=\int_0^t\frac{1}{N}\int_0^L\!\int_0^L c_h(x,y,t')\mathcal C_{\alpha,h}(x,y,t')\,\mathrm dx\,\mathrm dy\,\dd t',
 \qquad\alpha\in\{Q,g,a\},
 \label{eq:kernelcontributions}
\end{equation}
where $\mathcal C_{Q,h}=-2\langle\tau\rangle_hQ$, $\mathcal C_{g,h}=-\bfg\cdot\nabla\langle\tau\rangle_h$, and $\mathcal C_{a,h}=\bfa\cdot\nabla\langle\tau\rangle_h$.
Accounting also for the small-inertia approximation and spatial averaging gives the relative-entropy budget
\begin{equation}
 \mathcal D_h(t)-\mathcal D_h(0)=\mathcal D_Q+\mathcal D_g+\mathcal D_a+\mathcal D_{\mathrm{rem}}+\mathcal D_{\mathrm{avg}}.
 \label{eq:kernelentropybudget}
\end{equation}
Here $\mathcal D_{\mathrm{rem}}$ is the entropy contribution from the difference between the actual droplet velocities and their small-inertia approximation [Eq.~\eqref{eq:slow}].
The contribution $\mathcal D_{\mathrm{avg}}$ accounts for spatial variations in fluid velocity and acceleration within the averaging kernel, which are not captured by their values at the kernel center (discussed in the final part of this section).

\begin{figure}[!t]
 \centering
 \input{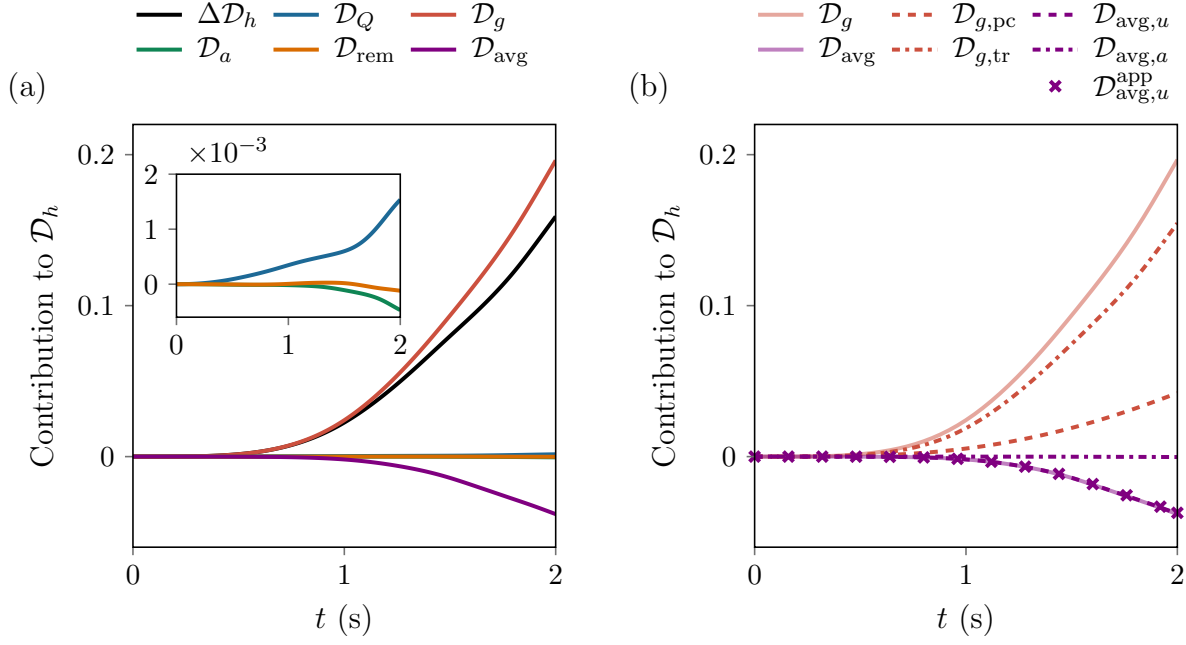}
 \setlength{\abovecaptionskip}{0pt}
 \caption{(a) Cumulative contributions to the relative-entropy increase of phase-changing droplets from strain and rotation ($\mathcal D_Q$), gravitational settling ($\mathcal D_g$), fluid acceleration ($\mathcal D_a$), the velocity remainder ($\mathcal D_{\mathrm{rem}}$), and spatial averaging ($\mathcal D_{\mathrm{avg}}$). Their sum gives $\Delta\mathcal D_h=\mathcal D_h(t)-\mathcal D_h(0)$ in Eq.~\eqref{eq:kernelentropybudget}. The inset enlarges $\mathcal D_Q$, $\mathcal D_a$, and $\mathcal D_{\mathrm{rem}}$ in units of $10^{-3}$. (b) Decomposition of the settling and spatial-averaging contributions. The settling contribution is reconstructed as the sum of its phase-change and transport history contributions, $\mathcal D_{g,\mathrm{pc}}$ and $\mathcal D_{g,\mathrm{tr}}$, in Eq.~\eqref{eq:settlinghistoryintegrals}. The spatial-averaging contribution is separated into velocity and acceleration contributions, $\mathcal D_{\mathrm{avg},u}$ and $\mathcal D_{\mathrm{avg},a}$. The approximation $\mathcal D_{\mathrm{avg},u}^{\mathrm{app}}$ is the time integral of Eq.~\eqref{eq:fluidaveragingstrain}.}
 \label{fig:countbudgethistory}
\end{figure}

Figure~\ref{fig:countbudgethistory}(a) shows the cumulative contributions to the relative-entropy increase.
At $t=2\,\mathrm s$, $\mathcal D_g=0.1962$, about $123\%$ of the relative-entropy increase for phase-changing droplets.
This contribution is partly offset by spatial averaging, with $\mathcal D_{\mathrm{avg}}=-0.03810$.
By contrast, the contribution from local strain and rotation, $\mathcal D_Q=0.001536$, is less than $1\%$ of this entropy increase, consistent with the weak preferential concentration of fixed-radius particles in Fig.~\ref{fig:hrbgeometry}.
The remaining contributions, $\mathcal D_a$ and $\mathcal D_{\mathrm{rem}}$, are both negative and smaller in magnitude than $\mathcal D_Q$.
The sum of the five contributions, $0.1590$, agrees with the relative-entropy increase of $0.1589$ calculated directly from Eq.~\eqref{eq:concentrationentropy}.
These results reveal that spatial differences in settling speed generated by phase change dominate the enhancement of preferential concentration.

\emph{Contributions of thermodynamic history.}
To explain how the settling contribution accumulates, we examine how phase change and droplet transport alter its rate, $\dot{\mathcal D}_g$.
Differentiating Eq.~\eqref{eq:kernelcontributions} twice in time gives
\begin{equation}
 \ddot{\mathcal D}_g
 =\frac{1}{N}\int_0^L\!\int_0^L
 \left[c_h\partial_t\mathcal C_{g,h}
 +\mathcal C_{g,h}\partial_t c_h\right]\,\mathrm dx\,\mathrm dy.
 \label{eq:euleriansettlingrate}
\end{equation}
The concentration satisfies $\partial_t c_h=-\nabla\cdot(c_h\langle\bfv\rangle_h)$, and the local mean response time satisfies $\partial_t\langle\tau\rangle_h=\langle\dot\tau\rangle_h-c_h^{-1}[\nabla\cdot(c_h\langle\tau\bfv\rangle_h)-\langle\tau\rangle_h\nabla\cdot(c_h\langle\bfv\rangle_h)]$.
We substitute these equations into Eq.~\eqref{eq:euleriansettlingrate}, using $\partial_t\mathcal C_{g,h}=-\bfg\cdot\nabla\partial_t\langle\tau\rangle_h$.
Integration by parts over the periodic domain then separates $\ddot{\mathcal D}_g$ into the phase-change (pc) and transport (tr) contributions:
\begin{align}
 \ddot{\mathcal D}_{g,\mathrm{pc}}
 &=\frac{1}{N}\int_0^L\!\int_0^L
 \langle\dot\tau\rangle_h\bfg\cdot\nabla c_h\,\mathrm dx\,\mathrm dy,
 \label{eq:entropyphasesource}\\
 \ddot{\mathcal D}_{g,\mathrm{tr}}
 &=\frac{1}{N}\int_0^L\!\int_0^L
 \frac{\nabla\cdot(c_h\langle\bfv\rangle_h)\,\bfg\cdot\nabla(c_h\langle\tau\rangle_h)
 -\nabla\cdot(c_h\langle\tau\bfv\rangle_h)\,\bfg\cdot\nabla c_h}{c_h}
 \,\mathrm dx\,\mathrm dy.
 \label{eq:entropytransportsource}
\end{align}
The transport contribution accounts for the transport of droplet number and response time through the fluxes $c_h\langle\bfv\rangle_h$ and $c_h\langle\tau\bfv\rangle_h$, respectively.
The initially uniform response time gives $\dot{\mathcal D}_g(0)=0$.
Integrating twice therefore gives
\begin{equation}
 \mathcal D_g(t)
 =\underbrace{\int_0^t(t-t')\ddot{\mathcal D}_{g,\mathrm{pc}}(t')\,\dd t'}_{\mathcal D_{g,\mathrm{pc}}(t)}
 +\underbrace{\int_0^t(t-t')\ddot{\mathcal D}_{g,\mathrm{tr}}(t')\,\dd t'}_{\mathcal D_{g,\mathrm{tr}}(t)}.
 \label{eq:settlinghistoryintegrals}
\end{equation}
The factor $t-t'$ accounts for the time over which a change in $\dot{\mathcal D}_g$ contributes to $\mathcal D_g$.
Detailed derivations of the moment equations and the settling-history decomposition above are given in Secs.~I~C--I~D of the Supplemental Material.

Figure~\ref{fig:countbudgethistory}(b) shows the cumulative phase-change and transport contributions to $\mathcal D_g$, together with their sum.
At $t=2\,\mathrm s$, these contributions are $\mathcal D_{g,\mathrm{pc}}=0.04209$ and $\mathcal D_{g,\mathrm{tr}}=0.15456$.
Their sum, $0.19665$, closely agrees with $\mathcal D_g=0.19620$ evaluated directly from Eq.~\eqref{eq:kernelcontributions}.
Notably, droplet transport accounts for about $79\%$ of this sum, substantially exceeding the direct contribution from condensation and evaporation.
Thus, although phase change creates the droplet size differences, droplet transport substantially amplifies the settling contribution by rearranging these droplets.

\emph{Fluid advection and concentration structure.}
Having identified the dominant settling contribution and traced its origins to phase change and droplet transport, we now examine the substantial negative contribution $\mathcal D_{\mathrm{avg}}$ in Fig.~\ref{fig:countbudgethistory}(a).
At $t=2\,\mathrm s$, it offsets about $19\%$ of the settling contribution.
To understand its physical origin, we separate $\mathcal D_{\mathrm{avg}}=\mathcal D_{\mathrm{avg},u}+\mathcal D_{\mathrm{avg},a}$ into contributions from spatial variations in fluid velocity and the inertial acceleration correction, respectively.
Their rates are
\begin{align}
 \dot{\mathcal D}_{\mathrm{avg},u}
 &=\frac1N\int_0^L\!\int_0^L
 \sum_p[\bfu(\bfx_p,t)-\bfu(\bfx,t)]W_h(\bfx-\bfx_p)
 \cdot\nabla\ln c_h\,\mathrm dx\,\mathrm dy,
 \label{eq:fluidaveragingrate}\\
 \dot{\mathcal D}_{\mathrm{avg},a}
 &=-\frac1N\int_0^L\!\int_0^L
 \sum_p\tau_p[\bfa(\bfx_p,t)-\bfa(\bfx,t)]W_h(\bfx-\bfx_p)
 \cdot\nabla\ln c_h\,\mathrm dx\,\mathrm dy.
 \label{eq:accelerationaveragingrate}
\end{align}
Figure~\ref{fig:countbudgethistory}(b) shows the cumulative velocity and acceleration contributions.
At $t=2\,\mathrm s$, these are $\mathcal D_{\mathrm{avg},u}=-0.03781$ and $\mathcal D_{\mathrm{avg},a}=-0.0002925$.
The velocity contribution accounts for $99.2\%$ of $\mathcal D_{\mathrm{avg}}$, while the acceleration contribution is negligibly small.

\begin{figure}[!t]
 \centering
 \input{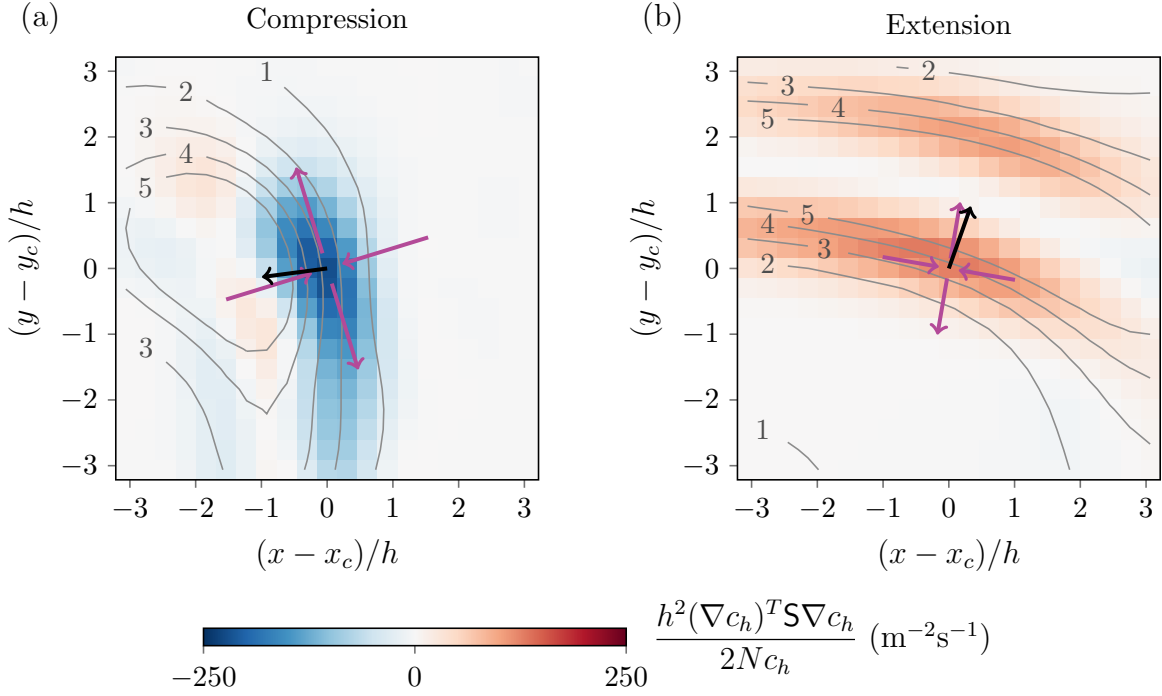}
 \setlength{\abovecaptionskip}{0pt}
 \caption{(a) Compression and (b) extension along the concentration gradient in two local regions at $t=2\,\mathrm s$. Colors show $h^2(\nabla c_h)^T\mathsf S\nabla c_h/(2Nc_h)$, whose spatial integral gives $\dot{\mathcal D}_{\mathrm{avg},u}^{\mathrm{app}}$ in Eq.~\eqref{eq:fluidaveragingstrain}. Gray contours show the normalized concentration $c_h/(N/L^2)$, and the black arrow indicates the direction of its gradient at the region center. Purple arrows indicate the principal axes of the strain-rate tensor $\mathsf S$ at the region center: inward arrows denote compression and outward arrows denote extension. Their lengths are proportional to the eigenvalue magnitudes, with the same scale in both panels. Coordinates are relative to the region center $\bfx_c=(x_c,y_c)$ and normalized by $h=1.276\,\mathrm{mm}$.}
 \label{fig:fluidadvection}
\end{figure}

We therefore expand Eq.~\eqref{eq:fluidaveragingrate} for small kernel width $h$.
For smooth velocity and concentration fields and the adopted quintic B-spline kernel $W_h$, this gives
\begin{equation}
 \dot{\mathcal D}_{\mathrm{avg},u}
 \simeq\dot{\mathcal D}_{\mathrm{avg},u}^{\mathrm{app}}\equiv\frac{h^2}{2N}\int_0^L\!\int_0^L
 \frac{(\nabla c_h)^T\mathsf S\nabla c_h}{c_h}\,\mathrm dx\,\mathrm dy.
 \label{eq:fluidaveragingstrain}
\end{equation}
The time-integrated approximation closely reproduces the velocity contribution in Fig.~\ref{fig:countbudgethistory}(b), giving $\mathcal D_{\mathrm{avg},u}^{\mathrm{app}}=-0.03722$ at $t=2\,\mathrm s$, with a relative error of only $1.6\%$.
The sign of the integrand depends on the orientation of the concentration gradient relative to the principal axes of $\mathsf S$.
Compression along the gradient produces a negative contribution, whereas extension produces a positive one.
Figure~\ref{fig:fluidadvection} illustrates these two cases in our simulation.
The negative cumulative contribution in Fig.~\ref{fig:countbudgethistory}(b) implies that compression along concentration gradients makes a larger cumulative contribution in magnitude than extension.

This dependence on strain and concentration gradients also appears in analyses of the subgrid-scale scalar flux \cite{Aluie2022,Balarac2012}.
Incompressible advection stretches and thins concentration structures without changing the concentration within each fluid element \cite{Batchelor1959}.
As these structures become thinner, spatial averaging at a fixed scale $h$ smooths out concentration differences between high-concentration regions and their lower-concentration surroundings.
The dependence of particle concentration statistics on the observation scale was also demonstrated by \citet{Bec2007}.
Thus, although the classical $Q$ contribution is weak, fluid deformation substantially reduces the preferential concentration measured at scale $h$.

\section{Conclusions}
We have derived a small-inertia concentration law for cloud droplets undergoing phase change and gravitational settling.
The absence of a direct gravitational contribution at leading order \cite{Shaw2003,Bec2014,Gustavsson2014} relies on spatially uniform settling.
Condensation and evaporation break this uniformity by producing spatial differences in droplet size, making gravitational settling a leading-order mechanism for changes in droplet concentration.
In the two-dimensional homogeneous thermal convection case studied here, this mechanism dominates the enhancement of preferential concentration.

To quantify this enhancement, we spatially average the droplet concentration at a fixed spatial scale and calculate the relative entropy of the resulting concentration field.
Although phase change creates the droplet size differences, our results show that droplet transport increases the settling contribution more than condensation and evaporation do directly.
Moreover, fluid deformation can stretch and thin concentration structures, as in the incompressible advection of a passive scalar.
Averaging these thinner structures at the fixed spatial scale smooths out the differences between high- and low-concentration regions, thereby reducing the relative entropy.
Overall, we provide a theoretical framework for quantifying and interpreting the coupled effects of phase change and gravitational settling on the preferential concentration of cloud droplets.

\bibliography{references}
\clearpage
\def\combinedsupplement{}
\ifdefined\combinedsupplement
\setcounter{section}{0}
\setcounter{subsection}{0}
\setcounter{equation}{0}
\setcounter{figure}{0}
\setcounter{table}{0}
\renewcommand{\theHsection}{supp.\arabic{section}}
\renewcommand{\theHequation}{supp.\arabic{equation}}
\renewcommand{\theHfigure}{supp.\arabic{figure}}
\renewcommand{\theHtable}{supp.\arabic{table}}
\fi
\renewcommand{\theequation}{S\arabic{equation}}
\renewcommand{\thefigure}{S\arabic{figure}}
\renewcommand{\thetable}{S\arabic{table}}
\ifdefined\combinedsupplement
\phantomsection
\pdfbookmark[0]{Supplemental Material}{supplement.title}
\begin{center}
\bfseries Supplemental Material for ``Preferential concentration of cloud droplets: Coupled effects of phase change and gravitational settling''
\end{center}
\fi
\setcounter{tocdepth}{3}
\setcounter{secnumdepth}{3}
\ifdefined\combinedsupplement
\makeatletter
\begingroup
\let\suppstarttoc\@starttoc
\def\@starttoc#1{\suppstarttoc{stoc}}
\tableofcontents
\endgroup
\let\suppaddcontentsline\addcontentsline
\renewcommand{\addcontentsline}[3]{%
  \ifstrequal{#1}{toc}{\suppaddcontentsline{stoc}{#2}{#3}}{\suppaddcontentsline{#1}{#2}{#3}}}
\makeatother
\else
\tableofcontents
\fi
\clearpage

\section{Theoretical derivations}
\label{sec:theoreticalderivations}

\subsection{Small-inertia velocity expansion}
\label{sec:velocityexpansion}
We derive the velocity approximation used in the concentration law from the droplet equations
\begin{equation}
 \dot{\bm x}_p=\bm v_p,\qquad
 \dot{\bm v}_p=\frac{\bm u(\bm x_p,t)-\bm v_p}{\tau_p^{(\varepsilon)}}+\bm g,\qquad
 \dot{\bm\xi}_p=\bm{\mathcal R}(\bm\xi_p,\bm x_p,\bm v_p,t).
 \label{eq:suppparticle}
\end{equation}
Here $\bm u$ is the incompressible fluid velocity, $\bm x_p$ and $\bm v_p$ are the droplet position and velocity, and $\bm\xi_p$ is its internal state, whose evolution is specified by $\bm{\mathcal R}$.
We use
\begin{equation}
 \tau_p=\frac{2\rho_w r_p^2}{9\rho_a\nu},\qquad
 \tau_p^{(\varepsilon)}=\varepsilon\tau_p,\qquad
 \bm g=(1-\rho_a/\rho_w)\bm g_0,
\end{equation}
where $r_p$ is the radius, $\rho_w$ and $\rho_a$ are the water and air densities, $\nu$ is the air kinematic viscosity, and $\bm g_0$ is the constant gravitational acceleration.
We take $\varepsilon\to0$ on a fixed time interval, holding the flow and phase-change model fixed; $\varepsilon=1$ gives the physical equations.

For the integer-power expansion, we also neglect finite-Reynolds-number corrections to heat and mass transfer, setting $\mathrm{Sh}=\mathrm{Nu}=2$.
With quasi-steady droplet temperature, the phase-change model then gives \cite{Richter2021,MacMillan2022}
\begin{equation}
 \frac{\mathrm d r_p^2}{\mathrm dt}=\frac{2\rho_aD_v}{\rho_w}(q-q_*),\qquad
 \frac{c_{pa}}{\mathrm{Pr}}(T-T_p)+\frac{L_v}{\mathrm{Sc}}(q-q_*)=0.
 \label{eq:suppzeroreynolds}
\end{equation}
The ambient temperature and water-vapor mixing ratio are $T$ and $q$; $q_*=q_*(r_p,T_p,T)$ is the surface-equilibrium mixing ratio specified in Sec.~III of the main text.
The constants $D_v$, $c_{pa}$, and $L_v$ are vapor diffusivity, air specific heat at constant pressure, and latent heat of vaporization, with $\mathrm{Sc}=\nu/D_v$ and $\mathrm{Pr}=\nu/\kappa_T$, where $\kappa_T$ is air thermal diffusivity.
The heat balance determines $T_p$ from $r_p$ and the ambient conditions, so the evolving internal state is $\bm\xi_p=r_p^2$.
Its rate is independent of droplet velocity in this limit and is denoted by $\bm{\mathcal R}_0(\bm\xi,\bm x,t)$.

After the initial velocity transient, we seek the velocity as a smooth function $\bm V^{(\varepsilon)}(\bm x,\bm\xi,t)$ \cite{Maxey1987,Haller2008}.
The coordinates $\bm x$ and $\bm\xi$ are independent variables of $\bm V^{(\varepsilon)}(\bm x,\bm\xi,t)$.
Along a droplet trajectory, $\dot{\bm x}_p=\bm V^{(\varepsilon)}$ and $\dot{\bm\xi}_p=\bm{\mathcal R}_0$, so the chain rule gives
\begin{equation}
 \frac{\mathrm d\bm V^{(\varepsilon)}}{\mathrm dt}
 =\partial_t\bm V^{(\varepsilon)}
 +(\bm V^{(\varepsilon)}\cdot\nabla_x)\bm V^{(\varepsilon)}
 +(\bm{\mathcal R}_0\cdot\nabla_\xi)\bm V^{(\varepsilon)}.
\end{equation}
Substitution into the momentum equation yields
\begin{equation}
 \varepsilon\tau_p\left[\partial_t\bm V^{(\varepsilon)}
 +(\bm V^{(\varepsilon)}\cdot\nabla_x)\bm V^{(\varepsilon)}
 +(\bm{\mathcal R}_0\cdot\nabla_\xi)\bm V^{(\varepsilon)}\right]
 =\bm u-\bm V^{(\varepsilon)}+\varepsilon\tau_p\bm g.
 \label{eq:velocityinvariance}
\end{equation}
At fixed $(\bm x,\bm\xi,t)$, expand the velocity and its convective derivative as
\begin{align}
 \bm V^{(\varepsilon)}&\sim\sum_{n=0}^{\infty}\varepsilon^n\bm V_n(\bm x,\bm\xi,t),\label{eq:velocityseries}\\
 (\bm V^{(\varepsilon)}\cdot\nabla_x)\bm V^{(\varepsilon)}
 &\sim\sum_{n=0}^{\infty}\varepsilon^n\sum_{j=0}^n(\bm V_j\cdot\nabla_x)\bm V_{n-j}.
\end{align}
Matching successive powers of $\varepsilon$ in Eq.~\eqref{eq:velocityinvariance} gives \cite{BenderOrszag1999}
\begin{align}
 O(1):\quad &\bm V_0=\bm u,\\
 O(\varepsilon):\quad &\tau_p[\partial_t\bm u+(\bm u\cdot\nabla_x)\bm u]
 =-\bm V_1+\tau_p\bm g,\qquad
 \bm V_1=\tau_p(\bm g-\bm a),
\end{align}
where $\bm a=(\partial_t+\bm u\cdot\nabla_x)\bm u$ is the fluid acceleration and $\nabla_\xi\bm u=0$.
For $n\ge2$, the coefficient of $\varepsilon^n$ gives the recurrence relation
\begin{equation}
 \bm V_n=-\tau_p\left[(\partial_t+\bm{\mathcal R}_0\cdot\nabla_\xi)\bm V_{n-1}
 +\sum_{j=0}^{n-1}(\bm V_j\cdot\nabla_x)\bm V_{n-1-j}\right].
 \label{eq:velocityrecursion}
\end{equation}
Evaluating the first two coefficients at the evolving droplet state gives
\begin{equation}
 \bm v_p(t)=\bm u(\bm x_p(t),t)+\tau_p^{(\varepsilon)}(t)
 [\bm g-\bm a(\bm x_p(t),t)]+O(\varepsilon^2).
 \label{eq:derivedvelocity}
\end{equation}

To determine the initial transient and approximation error, define the velocity remainder
\begin{equation}
 \bm e_p=\bm v_p-\bm u-\tau_p^{(\varepsilon)}\bm b,\qquad \bm b=\bm g-\bm a.
 \label{eq:suppvelocityremainder}
\end{equation}
In the following calculation, fluid quantities are evaluated at $(\bm x_p(t),t)$, and $D_p=\partial_t+\bm v_p\cdot\nabla_x$ differentiates along the droplet trajectory.
Using $D_p\bm u=\bm a+[(\bm v_p-\bm u)\cdot\nabla_x]\bm u$ and Eq.~\eqref{eq:suppparticle}, we obtain
\begin{align}
 \dot{\bm e}_p
 &=\dot{\bm v}_p-D_p\bm u-\dot\tau_p^{(\varepsilon)}\bm b-\tau_p^{(\varepsilon)}D_p\bm b\nonumber\\
 &=-\frac{\bm e_p+\tau_p^{(\varepsilon)}\bm b}{\tau_p^{(\varepsilon)}}+\bm g-\bm a
 -[(\bm e_p+\tau_p^{(\varepsilon)}\bm b)\cdot\nabla_x]\bm u
 -\dot\tau_p^{(\varepsilon)}\bm b-\tau_p^{(\varepsilon)}D_p\bm b\nonumber\\
 &=-\frac{\bm e_p}{\tau_p^{(\varepsilon)}}-(\bm e_p\cdot\nabla_x)\bm u
 -\dot\tau_p^{(\varepsilon)}\bm b
 -\tau_p^{(\varepsilon)}[D_p\bm b+(\bm b\cdot\nabla_x)\bm u].
 \label{eq:slipremainder}
\end{align}
On the fixed interval, let $0<\tau_p^{(\varepsilon)}\le\varepsilon\tau_{\max}$ and $\|\nabla_x\bm u\|\le A_{\max}$.
Assume $\dot\tau_p^{(\varepsilon)}=O(\varepsilon)$ and bounded $\bm b$ and $D_p\bm b$.
The three terms $-\dot\tau_p^{(\varepsilon)}\bm b$, $-\tau_p^{(\varepsilon)}D_p\bm b$, and $-\tau_p^{(\varepsilon)}(\bm b\cdot\nabla_x)\bm u$ on the right-hand side of Eq.~\eqref{eq:slipremainder} then satisfy
\begin{equation}
 |\dot\tau_p^{(\varepsilon)}|\,|\bm b|
 +\tau_p^{(\varepsilon)}[|D_p\bm b|+\|\nabla_x\bm u\|\,|\bm b|]
 \le\varepsilon F_{\max}.
\end{equation}
Here $\tau_{\max}$, $A_{\max}$, and $F_{\max}$ are bounds independent of $\varepsilon$ and valid for all particles; the norms are Euclidean and induced matrix norms.
Taking the scalar product of Eq.~\eqref{eq:slipremainder} with $\bm e_p/|\bm e_p|$ gives
\begin{align}
 \frac{\mathrm d|\bm e_p|}{\mathrm dt}
 &\le-\frac{|\bm e_p|}{\tau_p^{(\varepsilon)}}
 +\|\nabla_x\bm u\|\,|\bm e_p|+\varepsilon F_{\max}\nonumber\\
 &\le-\lambda_\varepsilon|\bm e_p|+\varepsilon F_{\max},
 \qquad \lambda_\varepsilon=\frac1{\varepsilon\tau_{\max}}-A_{\max}>0.
\end{align}
Multiplying by $\exp[\lambda_\varepsilon(t-t_0)]$ and integrating from the initial time $t_0$ yields
\begin{align}
 |\bm e_p(t)|
 &\le|\bm e_p(t_0)|e^{-\lambda_\varepsilon(t-t_0)}
 +\varepsilon F_{\max}\int_{t_0}^t e^{-\lambda_\varepsilon(t-t')}\,\mathrm dt'\nonumber\\
 &=|\bm e_p(t_0)|e^{-\lambda_\varepsilon(t-t_0)}
 +\frac{\varepsilon F_{\max}}{\lambda_\varepsilon}
 [1-e^{-\lambda_\varepsilon(t-t_0)}].
 \label{eq:velocityremainderbound}
\end{align}
The first term is the decaying initial velocity error, and $\varepsilon F_{\max}/\lambda_\varepsilon=O(\varepsilon^2)$.
Once the initial term reaches this order, Eq.~\eqref{eq:derivedvelocity} holds.
The estimate also applies to velocity-dependent phase change under the same bounds.
For the physical system ($\varepsilon=1$), the approximation remains accurate when droplet response times are short compared with the timescales of flow variation and droplet-size evolution.
Our simulations satisfy these conditions, with an initial Stokes number $\mathrm{St}_{\eta,0}=\tau_p(0)/\tau_\eta(0)\simeq0.0101$, where $\tau_\eta$ is the viscous dissipation timescale, and direct checks of the velocity approximation error in Sec.~\ref{sec:IIverification}.

\subsection{Phase-space transport and moment equations}
\label{sec:concentrationderivation}
For a given flow realization, let $\bm z=(\bm x,\bm v,\bm\xi)$ denote the physical position, velocity, and internal-state coordinates.
The initial droplet number in a phase-space element is $f_0(\bm z_0)\,\mathrm d\bm z_0$.
The total number is $N=\int f_0(\bm z_0)\,\mathrm d\bm z_0$ and is constant on the interval considered.
If $\bm z_p(t;\bm z_0)$ is the trajectory starting from $\bm z_0$ at $t_0$ under Eq.~\eqref{eq:suppparticle}, the evolving distribution is
\begin{equation}
 f(\bm z,t)=\int f_0(\bm z_0)\,
 \delta[\bm z-\bm z_p(t;\bm z_0)]\,\mathrm d\bm z_0.
 \label{eq:populationtransport}
\end{equation}
The Dirac distribution places each initial contribution at its current state; the integral adds these contributions over the initial population.
Differentiating the Dirac distribution at fixed $\bm z_0$ gives
\begin{align}
 \partial_t\delta(\bm z-\bm z_p)
 &=-\dot{\bm x}_p\cdot\nabla_x\delta
 -\dot{\bm v}_p\cdot\nabla_v\delta-\dot{\bm\xi}_p\cdot\nabla_\xi\delta\nonumber\\
 &=-\nabla_x\cdot(\dot{\bm x}_p\delta)
 -\nabla_v\cdot(\dot{\bm v}_p\delta)-\nabla_\xi\cdot(\dot{\bm\xi}_p\delta).
\end{align}
Here the trajectory rates are functions of time, while the derivatives act on $\bm x$, $\bm v$, and $\bm\xi$.
The sampling property of the delta function and the particle equations give
\begin{align}
 \dot{\bm x}_p\delta&=\bm v\delta,\\
 \dot{\bm v}_p\delta&=\left[\frac{\bm u(\bm x,t)-\bm v}{\tau_p^{(\varepsilon)}(\bm\xi)}+\bm g\right]\delta,\\
 \dot{\bm\xi}_p\delta&=\bm{\mathcal R}(\bm\xi,\bm x,\bm v,t)\delta.
\end{align}
Substituting these identities inside the divergences and integrating over $\bm z_0$ gives the phase-space transport equation \cite{Fox2012,Subramaniam2020},
\begin{equation}
 \partial_t f+\nabla_x\cdot(\bm v f)
 +\nabla_v\cdot\left[\left(\frac{\bm u-\bm v}{\tau_p^{(\varepsilon)}}+\bm g\right)f\right]
 +\nabla_\xi\cdot(\bm{\mathcal R}f)=0.
 \label{eq:phasetransport}
\end{equation}

Define the number density and, where $c>0$, the local number-weighted average of a droplet quantity $\psi$ by
\begin{equation}
 c=\int f\,\mathrm d\bm v\,\mathrm d\bm\xi,\qquad
 \langle\psi\rangle=\frac1c\int\psi f\,\mathrm d\bm v\,\mathrm d\bm\xi.
 \label{eq:supplocalaverage}
\end{equation}
Assume that the number and response-time fluxes vanish at the boundaries of velocity and internal-state space.
Integrating Eq.~\eqref{eq:phasetransport} over these coordinates then gives
\begin{align}
 0&=\int\partial_t f\,\mathrm d\bm v\,\mathrm d\bm\xi
 +\int\nabla_x\cdot(\bm v f)\,\mathrm d\bm v\,\mathrm d\bm\xi\nonumber\\
 &=\partial_t c+\nabla_x\cdot\int\bm v f\,\mathrm d\bm v\,\mathrm d\bm\xi
 =\partial_t c+\nabla_x\cdot(c\langle\bm v\rangle).
 \label{eq:populationnumberbalance}
\end{align}
In particular, the local mean response time is
\begin{equation}
 \langle\tau\rangle=\frac1c\int\tau_p^{(\varepsilon)}(\bm\xi)f\,\mathrm d\bm v\,\mathrm d\bm\xi.
\end{equation}
Averaging Eq.~\eqref{eq:suppvelocityremainder} at fixed $\bm x$ yields
\begin{equation}
 \langle\bm v\rangle=\bm u+\langle\tau\rangle(\bm g-\bm a)+\langle\bm e\rangle.
 \label{eq:suppmeanvelocity}
\end{equation}
We assume that the first spatial derivatives of $\langle\tau\rangle$ remain $O(\varepsilon)$ and those of $\langle\bm e\rangle$ remain $O(\varepsilon^2)$, so that taking the divergence preserves the order of the velocity approximation.

Let $\mathsf A=\nabla_x\bm u$, with components $A_{ij}=\partial_j u_i$, and define the strain-rate and rotation-rate tensors by $\mathsf S=(\mathsf A+\mathsf A^T)/2$ and $\mathsf\Omega=(\mathsf A-\mathsf A^T)/2$.
We use the Einstein summation convention for repeated indices.
For incompressible flow, the second velocity-gradient invariant is $Q=(\Omega_{ij}\Omega_{ij}-S_{ij}S_{ij})/2$ \cite{Chong1990}.
The divergence of fluid acceleration satisfies \cite{Maxey1987}
\begin{equation}
 \nabla_x\cdot\bm a=-2Q.
 \label{eq:suppdivacceleration}
\end{equation}
Substituting Eq.~\eqref{eq:suppmeanvelocity} into the number balance gives the concentration law,
\begin{align}
 (\partial_t+\langle\bm v\rangle\cdot\nabla_x)\ln c
 &=-\nabla_x\cdot\langle\bm v\rangle\nonumber\\
 &=-\nabla_x\cdot\bm u-(\bm g-\bm a)\cdot\nabla_x\langle\tau\rangle
 -\langle\tau\rangle\nabla_x\cdot(\bm g-\bm a)-\nabla_x\cdot\langle\bm e\rangle\nonumber\\
 &=\underbrace{-2\langle\tau\rangle Q}_{\mathcal C_Q}
 \underbrace{-\bm g\cdot\nabla_x\langle\tau\rangle}_{\mathcal C_g}
 \underbrace{+\bm a\cdot\nabla_x\langle\tau\rangle}_{\mathcal C_a}+O(\varepsilon^2).
 \label{eq:suppconcentration}
\end{align}

To obtain the local mean response-time equation, multiply Eq.~\eqref{eq:phasetransport} by $\tau_p^{(\varepsilon)}(\bm\xi)$ and integrate.
At fixed phase-space coordinates, this weight is independent of $t$, $\bm x$, and $\bm v$; along a droplet trajectory it changes at the rate
\begin{equation}
 \dot\tau_p^{(\varepsilon)}=\bm{\mathcal R}\cdot\nabla_\xi\tau_p^{(\varepsilon)}.
\end{equation}
The time and spatial terms give
\begin{align}
 \int\tau_p^{(\varepsilon)}\partial_t f\,\mathrm d\bm v\,\mathrm d\bm\xi
 &=\partial_t\int\tau_p^{(\varepsilon)}f\,\mathrm d\bm v\,\mathrm d\bm\xi
 =\partial_t(c\langle\tau\rangle),\\
 \int\tau_p^{(\varepsilon)}\nabla_x\cdot(\bm v f)\,\mathrm d\bm v\,\mathrm d\bm\xi
 &=\nabla_x\cdot\int\tau_p^{(\varepsilon)}\bm v f\,\mathrm d\bm v\,\mathrm d\bm\xi
 =\nabla_x\cdot(c\langle\tau\bm v\rangle).
\end{align}
Integration by parts in velocity and internal state gives the other two terms:
\begin{align}
 &\int\tau_p^{(\varepsilon)}\nabla_v\cdot\left[\left(\frac{\bm u-\bm v}{\tau_p^{(\varepsilon)}}+\bm g\right)f\right]\,\mathrm d\bm v\,\mathrm d\bm\xi\nonumber\\
 &\qquad=-\int f\left(\frac{\bm u-\bm v}{\tau_p^{(\varepsilon)}}+\bm g\right)
 \cdot\nabla_v\tau_p^{(\varepsilon)}\,\mathrm d\bm v\,\mathrm d\bm\xi=0,\\
 &\int\tau_p^{(\varepsilon)}\nabla_\xi\cdot(\bm{\mathcal R}f)\,\mathrm d\bm v\,\mathrm d\bm\xi\nonumber\\
 &\qquad=-\int f\bm{\mathcal R}\cdot\nabla_\xi\tau_p^{(\varepsilon)}\,\mathrm d\bm v\,\mathrm d\bm\xi
 =-\int\dot\tau_p^{(\varepsilon)}f\,\mathrm d\bm v\,\mathrm d\bm\xi
 =-c\langle\dot\tau\rangle.
\end{align}
Combining them yields
\begin{equation}
 \partial_t(c\langle\tau\rangle)+\nabla_x\cdot(c\langle\tau\bm v\rangle)
 =c\langle\dot\tau\rangle.
 \label{eq:responsemomentbalance}
\end{equation}
The product rule and number conservation now give
\begin{align}
 c\partial_t\langle\tau\rangle
 &=\partial_t(c\langle\tau\rangle)-\langle\tau\rangle\partial_t c\nonumber\\
 &=c\langle\dot\tau\rangle-\nabla_x\cdot(c\langle\tau\bm v\rangle)
 +\langle\tau\rangle\nabla_x\cdot(c\langle\bm v\rangle).
\end{align}
Thus the mean response time evolves according to
\begin{equation}
 \partial_t\langle\tau\rangle=\langle\dot\tau\rangle
 \underbrace{-\frac{\nabla_x\cdot(c\langle\tau\bm v\rangle)
 -\langle\tau\rangle\nabla_x\cdot(c\langle\bm v\rangle)}{c}}_{\mathcal T_\tau}.
 \label{eq:meanresponsetransport}
\end{equation}
Here $\langle\dot\tau\rangle$ describes phase change and $\mathcal T_\tau$ describes droplet transport.

\subsection{Relative entropy and its contribution budget}
\label{sec:kernelentropyderivation}
The concentration law above assumes a smooth concentration field and smooth local averages.
To construct these fields from the finite set of discrete Lagrangian particles used in simulations, we introduce spatial averaging with a kernel.
We then use the resulting fields to calculate the relative entropy and derive its contribution budget.
Throughout the remaining derivations, $\varepsilon=1$, time starts at particle release, and the particle number $N$ and kernel width $h$ are constant.

On a periodic square of side $L$, we use the quintic B-spline kernel \cite{Price2012}
\begin{align}
 B_5(s)&=\frac{(3-|s|)_+^5-6(2-|s|)_+^5+15(1-|s|)_+^5}{120},\\
 W_h(x,y)&=\sum_{k,l\in\mathbb Z}\frac1{h^2}
 B_5\!\left(\frac{x+kL}{h}\right)B_5\!\left(\frac{y+lL}{h}\right),
 \label{eq:suppkernel}
\end{align}
where $(s)_+=\max(s,0)$ and $k,l$ run over integer periodic images.
The kernel is nonnegative and normalized, with $\int_0^L\!\int_0^L W_h\,\mathrm dx\,\mathrm dy=1$.
Define the concentration and local averages by
\begin{equation}
 c_h(\bm x,t)=\sum_p W_h(\bm x-\bm x_p(t)),\qquad
 \langle\psi\rangle_h(\bm x,t)=\frac{\sum_p\psi_p(t)W_h(\bm x-\bm x_p(t))}{c_h(\bm x,t)}.
 \label{eq:suppkernelmoments}
\end{equation}
Integrals involving these local averages are evaluated over $c_h>0$, with zero contribution elsewhere.
Differentiating the concentration gives
\begin{align}
 \partial_t c_h
 &=-\sum_p\bm v_p\cdot\nabla W_h(\bm x-\bm x_p)
 =-\nabla\cdot\sum_p\bm v_p W_h(\bm x-\bm x_p).
\end{align}
The sum is the particle number flux,
\begin{equation}
 \bm j_h=\sum_p\bm v_p W_h(\bm x-\bm x_p)=c_h\langle\bm v\rangle_h,
\end{equation}
so that
\begin{equation}
 \partial_t c_h+\nabla\cdot(c_h\langle\bm v\rangle_h)=0,\qquad
 \int_0^L\!\int_0^L c_h\,\mathrm dx\,\mathrm dy=N.
 \label{eq:kernelnumberbalance}
\end{equation}
For the response-time moment, the same calculation gives
\begin{align}
 \partial_t(c_h\langle\tau\rangle_h)
 &=\partial_t\sum_p\tau_p W_h(\bm x-\bm x_p)\nonumber\\
 &=\sum_p\dot\tau_p W_h-\sum_p\tau_p\bm v_p\cdot\nabla W_h\nonumber\\
 &=c_h\langle\dot\tau\rangle_h-\nabla\cdot(c_h\langle\tau\bm v\rangle_h).
 \label{eq:kernelresponsebalance}
\end{align}
These equations have the same form as Eqs.~\eqref{eq:populationnumberbalance} and \eqref{eq:responsemomentbalance}.
Combining them gives the evolution equation for the local mean response time,
\begin{align}
 \partial_t\langle\tau\rangle_h
 &=\frac{\partial_t(c_h\langle\tau\rangle_h)-\langle\tau\rangle_h\partial_t c_h}{c_h}\nonumber\\
 &=\langle\dot\tau\rangle_h-
 \frac{\nabla\cdot(c_h\langle\tau\bm v\rangle_h)-\langle\tau\rangle_h\nabla\cdot(c_h\langle\bm v\rangle_h)}{c_h}.
 \label{eq:kernelmeanresponse}
\end{align}

The relative entropy with respect to the uniform concentration $N/L^2$ is
\begin{equation}
 \mathcal D_h=\frac1N\int_0^L\!\int_0^L c_h\ln\!\left(\frac{L^2c_h}{N}\right)\,\mathrm dx\,\mathrm dy.
 \label{eq:budgetentropydefinition}
\end{equation}
At $c_h=0$, the integrand is defined as zero.
Using number conservation and integrating by parts over the periodic domain gives
\begin{align}
 \frac{\mathrm d\mathcal D_h}{\mathrm dt}
 &=\frac1N\int_0^L\!\int_0^L
 \left[1+\ln\!\left(\frac{L^2c_h}{N}\right)\right]\partial_t c_h\,\mathrm dx\,\mathrm dy\nonumber\\
 &=-\frac1N\int_0^L\!\int_0^L
 \ln\!\left(\frac{L^2c_h}{N}\right)\nabla\cdot\bm j_h\,\mathrm dx\,\mathrm dy\nonumber\\
 &=\frac1N\int_0^L\!\int_0^L\bm j_h\cdot\nabla\ln c_h\,\mathrm dx\,\mathrm dy\nonumber\\
 &=\frac1N\int_0^L\!\int_0^L\langle\bm v\rangle_h\cdot\nabla c_h\,\mathrm dx\,\mathrm dy\nonumber\\
 &=-\frac1N\int_0^L\!\int_0^L c_h\nabla\cdot\langle\bm v\rangle_h\,\mathrm dx\,\mathrm dy.
 \label{eq:kernelentropyflux}
\end{align}
Number conservation removes the constant term, and periodicity removes the boundary terms.

To connect this entropy rate to the concentration law, substitute Eq.~\eqref{eq:suppvelocityremainder} into $\bm j_h$ and add and subtract the fluid fields at the integration point $\bm x$:
\begin{align}
 \bm j_h
 &=\sum_p\{\bm u(\bm x_p,t)+\tau_p[\bm g-\bm a(\bm x_p,t)]+\bm e_p\}W_h(\bm x-\bm x_p)\nonumber\\
 &=\bm u(\bm x,t)\sum_pW_h+[\bm g-\bm a(\bm x,t)]\sum_p\tau_pW_h+\sum_p\bm e_pW_h\nonumber\\
 &\quad+\sum_p\{\bm u(\bm x_p,t)-\bm u(\bm x,t)
 -\tau_p[\bm a(\bm x_p,t)-\bm a(\bm x,t)]\}W_h\nonumber\\
 &=c_h[\bm u+\langle\tau\rangle_h(\bm g-\bm a)]+\bm j_{\mathrm{rem}}+\bm j_{\mathrm{avg}},
 \label{eq:budgetfluxsplit}
\end{align}
where the last two fluxes are
\begin{align}
 \bm j_{\mathrm{rem}}&=\sum_p\bm e_pW_h(\bm x-\bm x_p),\label{eq:budgetremainderflux}\\
 \bm j_{\mathrm{avg}}&=\sum_p\{\bm u(\bm x_p,t)-\bm u(\bm x,t)
 -\tau_p[\bm a(\bm x_p,t)-\bm a(\bm x,t)]\}W_h(\bm x-\bm x_p).
 \label{eq:budgetaveragingflux}
\end{align}
They account for the velocity remainder and for spatial variations in fluid velocity and acceleration across the kernel, respectively.
Constant gravity has no averaging correction because $\sum_p\tau_p\bm gW_h=c_h\langle\tau\rangle_h\bm g$.

For the first flux in Eq.~\eqref{eq:budgetfluxsplit}, integration by parts and Eq.~\eqref{eq:suppdivacceleration} give
\begin{align}
 &\frac1N\int_0^L\!\int_0^L c_h[\bm u+\langle\tau\rangle_h(\bm g-\bm a)]\cdot\nabla\ln c_h\,\mathrm dx\,\mathrm dy\nonumber\\
 &\quad=-\frac1N\int_0^L\!\int_0^L c_h\nabla\cdot[\bm u+\langle\tau\rangle_h(\bm g-\bm a)]\,\mathrm dx\,\mathrm dy\nonumber\\
 &\quad=\frac1N\int_0^L\!\int_0^L c_h
 [-2\langle\tau\rangle_hQ-\bm g\cdot\nabla\langle\tau\rangle_h+\bm a\cdot\nabla\langle\tau\rangle_h]\,\mathrm dx\,\mathrm dy.
\end{align}
The three terms define the concentration-law contributions,
\begin{align}
 \dot{\mathcal D}_Q&=-\frac2N\int_0^L\!\int_0^L c_h\langle\tau\rangle_hQ\,\mathrm dx\,\mathrm dy,\\
 \dot{\mathcal D}_g&=-\frac1N\int_0^L\!\int_0^L c_h\bm g\cdot\nabla\langle\tau\rangle_h\,\mathrm dx\,\mathrm dy,\label{eq:suppsettlingrate}\\
 \dot{\mathcal D}_a&=\frac1N\int_0^L\!\int_0^L c_h\bm a\cdot\nabla\langle\tau\rangle_h\,\mathrm dx\,\mathrm dy.
\end{align}
The other two fluxes contribute
\begin{align}
 \dot{\mathcal D}_{\mathrm{rem}}&=\frac1N\int_0^L\!\int_0^L\bm j_{\mathrm{rem}}\cdot\nabla\ln c_h\,\mathrm dx\,\mathrm dy,\\
 \dot{\mathcal D}_{\mathrm{avg}}&=\frac1N\int_0^L\!\int_0^L\bm j_{\mathrm{avg}}\cdot\nabla\ln c_h\,\mathrm dx\,\mathrm dy.
 \label{eq:suppaveragingrate}
\end{align}
Together they give the exact rate budget
\begin{equation}
 \frac{\mathrm d\mathcal D_h}{\mathrm dt}
 =\dot{\mathcal D}_Q+\dot{\mathcal D}_g+\dot{\mathcal D}_a+\dot{\mathcal D}_{\mathrm{rem}}+\dot{\mathcal D}_{\mathrm{avg}}.
\end{equation}
Define each cumulative contribution from release by
\begin{equation}
 \mathcal D_\alpha(t)=\int_0^t\dot{\mathcal D}_\alpha(t')\,\mathrm dt',
 \qquad \alpha\in\{Q,g,a,\mathrm{rem},\mathrm{avg}\}.
\end{equation}
Integrating the rate budget gives
\begin{equation}
 \mathcal D_h(t)-\mathcal D_h(0)
 =\mathcal D_Q+\mathcal D_g+\mathcal D_a+\mathcal D_{\mathrm{rem}}+\mathcal D_{\mathrm{avg}}.
 \label{eq:budgetcumulativeidentity}
\end{equation}

\subsection{History decomposition of the settling contribution}
\label{sec:settlinghistory}
We use the moment balances to determine how phase change and droplet transport alter the settling contribution.
Differentiating Eq.~\eqref{eq:suppsettlingrate} gives
\begin{equation}
 N\ddot{\mathcal D}_g=-\int_0^L\!\int_0^L
 [c_h\bm g\cdot\nabla\partial_t\langle\tau\rangle_h
 +(\partial_t c_h)\bm g\cdot\nabla\langle\tau\rangle_h]\,\mathrm dx\,\mathrm dy.
 \label{eq:suppsettlingsecondrate}
\end{equation}
Substitute Eqs.~\eqref{eq:kernelnumberbalance} and \eqref{eq:kernelmeanresponse}, and integrate the first term by parts:
\begin{align}
 N\ddot{\mathcal D}_g
 &=\int_0^L\!\int_0^L
 \left[\partial_t\langle\tau\rangle_h\,\bm g\cdot\nabla c_h
 +\nabla\cdot(c_h\langle\bm v\rangle_h)\,\bm g\cdot\nabla\langle\tau\rangle_h\right]\mathrm dx\,\mathrm dy\nonumber\\
 &=\int_0^L\!\int_0^L\left[
 \langle\dot\tau\rangle_h\bm g\cdot\nabla c_h
 -\frac{\nabla\cdot(c_h\langle\tau\bm v\rangle_h)}{c_h}\bm g\cdot\nabla c_h\right.\nonumber\\
 &\qquad\left.+\frac{\langle\tau\rangle_h\nabla\cdot(c_h\langle\bm v\rangle_h)}{c_h}\bm g\cdot\nabla c_h
 +\nabla\cdot(c_h\langle\bm v\rangle_h)\,\bm g\cdot\nabla\langle\tau\rangle_h
 \right]\mathrm dx\,\mathrm dy\nonumber\\
 &=\int_0^L\!\int_0^L\left[
 \langle\dot\tau\rangle_h\bm g\cdot\nabla c_h
 -\frac{\nabla\cdot(c_h\langle\tau\bm v\rangle_h)}{c_h}\bm g\cdot\nabla c_h+\frac{\langle\tau\rangle_h\nabla\cdot(c_h\langle\bm v\rangle_h)}{c_h}\bm g\cdot\nabla c_h\right.\nonumber\\
 &\qquad\left.+\nabla\cdot(c_h\langle\bm v\rangle_h)
 \left(\frac{\bm g\cdot\nabla(c_h\langle\tau\rangle_h)}{c_h}
 -\frac{\langle\tau\rangle_h}{c_h}\bm g\cdot\nabla c_h\right)
 \right]\mathrm dx\,\mathrm dy\nonumber\\
 &=\int_0^L\!\int_0^L\left[
 \langle\dot\tau\rangle_h\bm g\cdot\nabla c_h+\frac{\nabla\cdot(c_h\langle\bm v\rangle_h)\,\bm g\cdot\nabla(c_h\langle\tau\rangle_h)
 -\nabla\cdot(c_h\langle\tau\bm v\rangle_h)\,\bm g\cdot\nabla c_h}{c_h}
 \right]\mathrm dx\,\mathrm dy.
\end{align}
Thus the phase-change (pc) and transport (tr) contributions are
\begin{align}
 \ddot{\mathcal D}_g&=\ddot{\mathcal D}_{g,\mathrm{pc}}+\ddot{\mathcal D}_{g,\mathrm{tr}},\\
 \ddot{\mathcal D}_{g,\mathrm{pc}}
 &=\frac1N\int_0^L\!\int_0^L\langle\dot\tau\rangle_h\bm g\cdot\nabla c_h\,\mathrm dx\,\mathrm dy,\label{eq:kernelhistoryphaseevaluation}\\
 \ddot{\mathcal D}_{g,\mathrm{tr}}
 &=\frac1N\int_0^L\!\int_0^L
 \frac{\nabla\cdot(c_h\langle\bm v\rangle_h)\,\bm g\cdot\nabla(c_h\langle\tau\rangle_h)
 -\nabla\cdot(c_h\langle\tau\bm v\rangle_h)\,\bm g\cdot\nabla c_h}{c_h}\,\mathrm dx\,\mathrm dy.
 \label{eq:kernelhistorytransportevaluation}
\end{align}
In $\ddot{\mathcal D}_{g,\mathrm{tr}}$, the flux $c_h\langle\bm v\rangle_h$ describes the transport of droplet number, while $c_h\langle\tau\bm v\rangle_h$ describes the transport of droplet response time.

Initially equal response times imply $\langle\tau\rangle_h(\bm x,0)=\tau_p(0)$ and $\dot{\mathcal D}_g(0)=0$.
With $\mathcal D_g(0)=0$, integrating twice gives
\begin{equation}
 \mathcal D_g(t)
 =\underbrace{\int_0^t(t-t')\ddot{\mathcal D}_{g,\mathrm{pc}}(t')\,\mathrm dt'}_{\mathcal D_{g,\mathrm{pc}}(t)}
 +\underbrace{\int_0^t(t-t')\ddot{\mathcal D}_{g,\mathrm{tr}}(t')\,\mathrm dt'}_{\mathcal D_{g,\mathrm{tr}}(t)}.
 \label{eq:supphistoryintegrals}
\end{equation}

\subsection{Spatial-averaging contributions}
\label{sec:averagingphysics}
We separate the velocity and acceleration differences in Eq.~\eqref{eq:budgetaveragingflux} to obtain
\begin{align}
 \bm j_{\mathrm{avg}}&=\bm j_{\mathrm{avg},u}+\bm j_{\mathrm{avg},a},\\
 \bm j_{\mathrm{avg},u}&=\sum_p[\bm u(\bm x_p,t)-\bm u(\bm x,t)]W_h(\bm x-\bm x_p),\\
 \bm j_{\mathrm{avg},a}&=-\sum_p\tau_p[\bm a(\bm x_p,t)-\bm a(\bm x,t)]W_h(\bm x-\bm x_p).
 \label{eq:averagingfluxparts}
\end{align}
Substitution into Eq.~\eqref{eq:suppaveragingrate} gives
\begin{align}
 \dot{\mathcal D}_{\mathrm{avg},u}&=\frac1N\int_0^L\!\int_0^L\bm j_{\mathrm{avg},u}\cdot\nabla\ln c_h\,\mathrm dx\,\mathrm dy,\\
 \dot{\mathcal D}_{\mathrm{avg},a}&=\frac1N\int_0^L\!\int_0^L\bm j_{\mathrm{avg},a}\cdot\nabla\ln c_h\,\mathrm dx\,\mathrm dy.
 \label{eq:averagingrateparts}
\end{align}
Integrating each rate from $0$ to $t$ defines $\mathcal D_{\mathrm{avg},u}$ and $\mathcal D_{\mathrm{avg},a}$, with $\mathcal D_{\mathrm{avg}}=\mathcal D_{\mathrm{avg},u}+\mathcal D_{\mathrm{avg},a}$.
The contribution from the fluid velocity evaluated at $\bm x$ vanishes:
\begin{equation}
 \int_0^L\!\int_0^L c_h\bm u\cdot\nabla\ln c_h\,\mathrm dx\,\mathrm dy
 =\int_0^L\!\int_0^L\bm u\cdot\nabla c_h\,\mathrm dx\,\mathrm dy
 =-\int_0^L\!\int_0^L c_h\nabla\cdot\bm u\,\mathrm dx\,\mathrm dy=0.
\end{equation}

To approximate $\dot{\mathcal D}_{\mathrm{avg},u}$ for small $h$, we consider an idealized smooth, strictly positive periodic concentration $c$, defined in Eq.~\eqref{eq:supplocalaverage} before spatial averaging, and a smooth periodic fluid velocity $\bm u$.
When sufficiently many particles sample this continuous distribution, the kernel sums can be approximated by integrals weighted by $c$.
For a periodic scalar or vector field $F$, define the convolution by
\begin{equation}
 (W_h*F)(x,y,t)=\int_0^L\!\int_0^L W_h(x-x',y-y')F(x',y',t)\,\mathrm dx'\,\mathrm dy'.
\end{equation}
In this continuum approximation, the concentration and velocity flux are
\begin{equation}
 c_h=W_h*c,\qquad \bm j_{\mathrm{avg},u}=W_h*(c\bm u)-\bm u(W_h*c).
 \label{eq:averagingcontinuumfluxes}
\end{equation}
To expand the convolution in the displacement $\bm\eta=\bm x-\bm x'$ from the source point $\bm x'$ to the evaluation point $\bm x$, we write it equivalently as
\begin{equation}
 (W_h*F)(\bm x,t)=\int_{\mathbb R^2}w_h(\bm\eta)F(\bm x-\bm\eta,t)\,\mathrm d\eta_x\,\mathrm d\eta_y.
\end{equation}
Here $w_h(\bm\eta)=h^{-2}B_5(\eta_x/h)B_5(\eta_y/h)$, and values of $F$ outside the domain are evaluated using its periodicity.
The normalization and second moment of $B_5$ follow directly from its definition:
\begin{align}
 \int_{-\infty}^{\infty}B_5(s)\,\mathrm ds
 &=\frac1{60}\left(\frac{3^6}{6}-6\frac{2^6}{6}+\frac{15}{6}\right)=1,\\
 \int_{-\infty}^{\infty}s^2B_5(s)\,\mathrm ds
 &=\frac1{60}\left(\frac{3^8}{168}-6\frac{2^8}{168}+\frac{15}{168}\right)=\frac12.
\end{align}
Symmetry removes odd moments, while scaling by $h$ gives
\begin{equation}
 \int_{\mathbb R^2}\eta_i\eta_jw_h\,\mathrm d\eta_x\,\mathrm d\eta_y
 =\frac{h^2}{2}\delta_{ij},
\end{equation}
where $\delta_{ij}$ is the Kronecker delta.
Taylor expansion of $F(\bm x-\bm\eta,t)$ about $\bm x$ and termwise integration give
\begin{align}
 (W_h*F)(\bm x,t)
 &=\int_{\mathbb R^2}w_h(\bm\eta)F(\bm x-\bm\eta,t)\,\mathrm d\eta_x\,\mathrm d\eta_y\nonumber\\
 &=\int_{\mathbb R^2}w_h(\bm\eta)\left[F(\bm x,t)-\eta_i\partial_iF(\bm x,t)
 +\frac12\eta_i\eta_j\partial_i\partial_jF(\bm x,t)\right.\nonumber\\
 &\qquad\left.-\frac16\eta_i\eta_j\eta_k\partial_i\partial_j\partial_kF(\bm x,t)
 +O(|\bm\eta|^4)\right]\mathrm d\eta_x\,\mathrm d\eta_y\nonumber\\
 &=F(\bm x,t)+\frac{h^2}{4}\nabla^2F(\bm x,t)+O(h^4).
 \label{eq:averagingtaylorexpansion}
\end{align}
Applying this result to both convolutions in Eq.~\eqref{eq:averagingcontinuumfluxes} yields
\begin{align}
 W_h*(c\bm u)&=c\bm u+\frac{h^2}{4}\nabla^2(c\bm u)+O(h^4),\\
 \bm u(W_h*c)&=c\bm u+\frac{h^2}{4}\bm u\nabla^2c+O(h^4).
\end{align}
Subtracting the two convolutions therefore gives
\begin{equation}
 \bm j_{\mathrm{avg},u}=\frac{h^2}{2}(\nabla c\cdot\nabla)\bm u+\frac{h^2}{4}c\nabla^2\bm u+O(h^4).
 \label{eq:averagingvelocityexpansion}
\end{equation}
Expanding the concentration and its gradient gives $c_h=c+O(h^2)$ and $\nabla c_h=\nabla c+O(h^2)$, so
\begin{equation}
 \nabla\ln c_h=\frac{\nabla c_h}{c_h}=\frac{\nabla c}{c}+O(h^2).
\end{equation}
Substituting these two expressions into $\dot{\mathcal D}_{\mathrm{avg},u}$ gives
\begin{align}
 \dot{\mathcal D}_{\mathrm{avg},u}
 &=\frac1N\int_0^L\!\int_0^L
 \left[\frac{h^2}{2}(\nabla c\cdot\nabla)\bm u+\frac{h^2}{4}c\nabla^2\bm u\right]\cdot\frac{\nabla c}{c}\,\mathrm dx\,\mathrm dy+O(h^4)\nonumber\\
 &=\frac{h^2}{2N}\int_0^L\!\int_0^L\frac{(\partial_i c)(\partial_j c)(\partial_j u_i)}{c}\,\mathrm dx\,\mathrm dy
 +\frac{h^2}{4N}\int_0^L\!\int_0^L\nabla^2\bm u\cdot\nabla c\,\mathrm dx\,\mathrm dy+O(h^4).
\end{align}
The second integral vanishes by incompressibility and periodicity:
\begin{equation}
 \int_0^L\!\int_0^L\nabla^2\bm u\cdot\nabla c\,\mathrm dx\,\mathrm dy
 =-\int_0^L\!\int_0^L c\nabla\cdot\nabla^2\bm u\,\mathrm dx\,\mathrm dy
 =-\int_0^L\!\int_0^L c\nabla^2(\nabla\cdot\bm u)\,\mathrm dx\,\mathrm dy=0.
\end{equation}
For the first integral, $\nabla\bm u=\mathsf S+\mathsf\Omega$ and the antisymmetric part contributes zero:
\begin{equation}
 (\nabla c)^T\mathsf\Omega\nabla c
 =[(\nabla c)^T\mathsf\Omega\nabla c]^T
 =(\nabla c)^T\mathsf\Omega^T\nabla c
 =-(\nabla c)^T\mathsf\Omega\nabla c=0.
\end{equation}
Consequently,
\begin{equation}
 \dot{\mathcal D}_{\mathrm{avg},u}
 =\frac{h^2}{2N}\int_0^L\!\int_0^L\frac{(\nabla c)^T\mathsf S\nabla c}{c}\,\mathrm dx\,\mathrm dy+O(h^4).
\end{equation}
Replacing $c$ and $\nabla c$ by their kernel averages changes the strain integral by $O(h^2)$ and hence the entropy rate by $O(h^4)$:
\begin{equation}
 \dot{\mathcal D}_{\mathrm{avg},u}
 =\underbrace{\frac{h^2}{2N}\int_0^L\!\int_0^L
 \frac{(\nabla c_h)^T\mathsf S\nabla c_h}{c_h}\,\mathrm dx\,\mathrm dy}_{\dot{\mathcal D}_{\mathrm{avg},u}^{\mathrm{app}}}+O(h^4).
 \label{eq:averagingresolvedstrainrate}
\end{equation}
For any normalized, symmetric kernel with an isotropic second moment, the leading-order expression has the same dependence on the strain-rate tensor and concentration gradient; the prefactor is determined by the kernel's second moment.
Compression along the concentration gradient makes a negative contribution; extension makes a positive one.

\section{Numerical implementation}
\label{sec:numericalimplementation}
\subsection{Physical models and parameters}
\label{sec:physicalmodel}
We simulate dilute water droplets in two-dimensional homogeneous thermal convection.
The fluid equations are
\begin{equation}
\begin{aligned}
 \partial_t\bm u+\bm u\cdot\nabla\bm u
 &=-\rho_a^{-1}\nabla p+\nu\nabla^2\bm u+\mu\nabla^{-2}\bm u+g_0\beta\theta\bm e_y,
 \qquad \nabla\cdot\bm u=0,\\
 \partial_t\theta+\bm u\cdot\nabla\theta
 &=\kappa_T\nabla^2\theta+(\Delta T/L)u_y.
\end{aligned}
\label{eq:IIfluid}
\end{equation}
Here $\bm u$ is the fluid velocity, $\theta$ the temperature fluctuation, $p$ the pressure, and $\bm e_y$ the upward unit vector.
The velocity and temperature fluctuation are periodic on a square of side $L$.
The background temperature gradient is $-\Delta T/L$.
The inverse-Laplacian term damps nonzero velocity modes at large scales, with $\mu=(2\pi)^{5/2}U_f/L^3$ and $U_f=(g_0\beta\Delta TL)^{1/2}$ \cite{Winchester2024}.
The zero velocity mode is set to zero.

The temperature sampled by a droplet is the sum of the linear background and the periodic fluctuation,
\begin{equation}
 T_p^{\mathrm{amb}}(t)=T_*+\Delta T/2-(\Delta T/L)y_p^{\mathrm{unwrapped}}(t)+\theta(\bm x_p(t),t),
 \qquad T_*=300\,\mathrm K.
 \label{eq:IIambient}
\end{equation}
The water-vapor mixing ratio is uniform and constant,
$q=M_we_s(T_*)/(R_u\rho_aT_*)$.
We use one-way coupling, so droplet motion and phase change do not feed back on the fluid velocity, temperature, or water-vapor mixing ratio.

For droplet mass $m_p=4\pi\rho_wr_p^3/3$, we solve
\begin{equation}
 \dot{\bm x}_p=\bm v_p,\qquad
 \dot{\bm v}_p=\frac{\bm u(\bm x_p,t)-\bm v_p}{\tau_p}+\bm g,\qquad
 \tau_p=\frac{2\rho_wr_p^2}{9\rho_a\nu},\qquad
 \bm g=-(1-\rho_a/\rho_w)g_0\bm e_y.
 \label{eq:IIdropletmotion}
\end{equation}
The pure-water mass-transfer model and quasi-steady heat balance are \cite{MacMillan2022,Richter2021,Vaillancourt2001}
\begin{equation}
 \dot m_p=2\pi r_p\rho_aD_v\mathrm{Sh}_p(q-q_{*,p}),\qquad
 0=2\pi r_p\rho_ac_{pa}\kappa_T\mathrm{Nu}_p(T_p^{\mathrm{amb}}-T_p)+L_v\dot m_p.
 \label{eq:IImassheat}
\end{equation}
Here $T_p$ is the droplet temperature, $D_v$ the vapor diffusivity, $c_{pa}$ the specific heat of air at constant pressure, and $L_v$ the latent heat of vaporization.
Using $\dot m_p=4\pi\rho_wr_p^2\dot r_p$, $\mathrm{Sc}=\nu/D_v$, and $\mathrm{Pr}=\nu/\kappa_T$ gives
\begin{align}
 \frac{\mathrm d r_p^2}{\mathrm dt}
 &=\frac{\rho_aD_v\mathrm{Sh}_p}{\rho_w}(q-q_{*,p}),
 \qquad\dot\tau_p=\frac{2\mathrm{Sh}_p}{9\mathrm{Sc}}(q-q_{*,p}),\label{eq:IIgrowth}\\
 0&=c_{pa}\frac{\mathrm{Nu}_p}{\mathrm{Pr}}(T_p^{\mathrm{amb}}-T_p)
 +L_v\frac{\mathrm{Sh}_p}{\mathrm{Sc}}(q-q_{*,p}).&&\label{eq:IIheat}
\end{align}
The surface value $q_{*,p}=q_*(r_p,T_p,T_p^{\mathrm{amb}})$ is evaluated from \cite{Richter2021}
\begin{equation}
 q_*(r,T_p,T)=\frac{M_we_s(T)}{R_uT_p\rho_a}
 \exp\left[\frac{L_vM_w}{R_u}\left(\frac1T-\frac1{T_p}\right)
 +\frac{2\sigma M_w}{R_u\rho_wrT_p}\right].
 \label{eq:IIsurface}
\end{equation}
The constants $M_w$, $R_u$, and $\sigma$ are the water molar mass, gas constant, and surface tension.
We use the modified Magnus formula~\cite{Alduchov1996} for the saturation vapor pressure over a plane water surface,
\begin{equation}
 e_s(T)=610.94\,\mathrm{Pa}\,
 \exp\left[\frac{17.625(T-273.15\,\mathrm K)}{243.04\,\mathrm K+T-273.15\,\mathrm K}\right].
\end{equation}
The Sherwood and Nusselt numbers use the Ranz--Marshall correlations \cite{Ranz1952I,Ranz1952II},
\begin{equation}
 \mathrm{Sh}_p=2+0.6\mathrm{Re}_p^{1/2}\mathrm{Sc}^{1/3},\qquad
 \mathrm{Nu}_p=2+0.6\mathrm{Re}_p^{1/2}\mathrm{Pr}^{1/3},\qquad
 \mathrm{Re}_p=\frac{2r_p|\bm v_p-\bm u(\bm x_p,t)|}{\nu}.
\end{equation}
These finite-Reynolds-number corrections are retained in the heat and mass transfer.
No activation, collision, or coalescence is included.

\begin{table}[!h]
\caption{Physical input parameters. Air and water properties follow MacMillan \emph{et al.}~\cite{MacMillan2022}.}
\begin{ruledtabular}
\begin{tabular}{lll}
Quantity & Value & Unit\\
$L$ & $0.20$ & m\\
$\Delta T$ & $4$ & K\\
$\rho_a$ & $1.2$ & kg m$^{-3}$\\
$\rho_w$ & $1000$ & kg m$^{-3}$\\
$\nu$ & $1.57\times10^{-5}$ & m$^2$ s$^{-1}$\\
$\mathrm{Pr}$ & $0.71$ & ---\\
$\mathrm{Sc}$ & $0.615$ & ---\\
$c_{pa}$ & $1006$ & J kg$^{-1}$ K$^{-1}$\\
$L_v$ & $2.44\times10^6$ & J kg$^{-1}$\\
$\sigma$ & $0.0728$ & N m$^{-1}$\\
$M_w$ & $0.018$ & kg mol$^{-1}$\\
$R_u$ & $8.314$ & J mol$^{-1}$ K$^{-1}$\\
$g_0$ & $9.81$ & m s$^{-2}$\\
$\beta$ & $1/300$ & K$^{-1}$\\
$r_0$ & $15$ & $\mu$m\\
\end{tabular}
\end{ruledtabular}
\label{tab:IIphysical}
\end{table}
The parameters in Table~\ref{tab:IIphysical} give $\kappa_T=2.211\times10^{-5}\,\mathrm{m^2\,s^{-1}}$, $D_v=2.553\times10^{-5}\,\mathrm{m^2\,s^{-1}}$, $q=0.02122$, and $\mu=2001\,\mathrm{m^{-2}\,s^{-1}}$.
The free-fall time is $t_f=L/U_f=1.237\,\mathrm s$, the Rayleigh number is $Ra=g_0\beta\Delta TL^3/(\nu\kappa_T)=3.01\times10^6$, and the initial response time is $2.654\,\mathrm{ms}$.

\subsection{Numerical methods}
\label{sec:IImethods}
\subsubsection{Flow solver}
The fluid equations are solved using a pseudospectral method with $128^2$ Fourier modes.
Nonlinear products are evaluated on a $192^2$ physical grid obtained by zero-padding the Fourier coefficients and applying the inverse Fourier transform.
The products are then transformed back to Fourier space and truncated to the retained modes, implementing the $3/2$ dealiasing rule \cite{Peyret2002}.
Nyquist modes are removed.
Time integration uses a third-order integrating-factor Runge--Kutta method \cite{Lawson1967}, which treats viscosity, thermal diffusion, and inverse-Laplacian damping exactly.
Let $Y$ collect the Fourier coefficients of $\bm u$ and $\theta$, so that $Y'=\mathcal LY+\mathcal N(Y)$.
For each nonzero wavevector $\bm k=2\pi(m,n)/L$, with integers $m,n$, the two operators are
\begin{equation}
 (\mathcal LY)_{\bm k}
 =\begin{pmatrix}
 -\bigl(\nu|\bm k|^2+\mu/|\bm k|^2\bigr)\widehat{\bm u}_{\bm k}\\
 -\kappa_T|\bm k|^2\widehat\theta_{\bm k}
 \end{pmatrix},\qquad
 [\mathcal N(Y)]_{\bm k}
 =\begin{pmatrix}
 \mathsf P_{\bm k}\bigl[-\widehat{(\bm u\cdot\nabla)\bm u}_{\bm k}
 +g_0\beta\widehat\theta_{\bm k}\bm e_y\bigr]\\
 -\widehat{\bm u\cdot\nabla\theta}_{\bm k}+(\Delta T/L)\widehat u_{y,\bm k}
 \end{pmatrix}.
\end{equation}
Here hats denote Fourier coefficients, and $\mathsf P_{\bm k}=\mathsf I-\bm k\bm k^T/|\bm k|^2$ enforces incompressibility, with $\mathsf I$ the identity tensor.
The integrating-factor transformation $Z(s)=e^{-\mathcal Ls}Y(t_n+s)$ gives $Z'=F(s,Z)$, where $F(s,Z)=e^{-\mathcal Ls}\mathcal N(e^{\mathcal Ls}Z)$.
Writing $Y_n=Y(t_n)$, the update over a time step $\Delta t$ is
\begin{align}
 K_1&=F(0,Y_n),\nonumber\\
 K_2&=F(\Delta t/2,Y_n+\Delta t K_1/2),\nonumber\\
 K_3&=F(\Delta t,Y_n-\Delta t K_1+2\Delta t K_2),\nonumber\\
 Y_{n+1}&=e^{\mathcal L\Delta t}\left[Y_n+\frac{\Delta t}{6}(K_1+4K_2+K_3)\right].
 \label{eq:IIrk}
\end{align}
The time step is $\Delta t=0.5\,\mathrm{ms}$.

We initialize the flow with a velocity perturbation specified below through its vorticity $\omega=\partial_xu_y-\partial_yu_x$ and set the temperature fluctuation to zero.
In the expansion $\omega=\sum_{m,n}\widehat\omega_{mn}e^{2\pi\mathrm i(mx+ny)/L}$, the nonzero coefficients with positive $n$ are
\begin{equation}
\begin{aligned}
 (\widehat\omega_{11},\widehat\omega_{21},\widehat\omega_{12})
 &=\frac{0.01}{t_f}(0.7-0.2\mathrm i,-0.3+0.5\mathrm i,0.4+0.6\mathrm i),\\
 (\widehat\omega_{31},\widehat\omega_{23},\widehat\omega_{32})
 &=\frac{0.01}{t_f}(-0.2-0.4\mathrm i,0.3-0.5\mathrm i,-0.6+0.1\mathrm i).
\end{aligned}
\end{equation}
The opposite-wavevector coefficients are their complex conjugates.

\subsubsection{Particle solver}
After the flow has evolved for $20t_f$, we release $512^2$ particles in each of the fixed-radius and phase-changing populations at $[(i+1/2)L/512,(j+1/2)L/512]$, with $i,j=0,\ldots,511$.
Both populations start with $r_0=15\,\mu\mathrm m$ and the interpolated fluid velocity.
Time in the particle results is measured from release.

We interpolate fluid quantities to particle positions using four-point Lagrange interpolation in each coordinate on the aforementioned $192^2$ physical grid.
For a particle at $(x_p,y_p)$ with $x_i\le x_p<x_{i+1}$ and $y_j\le y_p<y_{j+1}$, define its position within the grid cell by
\begin{equation}
 s_x=\frac{x_p-x_i}{\Delta x},\qquad s_y=\frac{y_p-y_j}{\Delta y},\qquad 0\le s_x,s_y<1.
\end{equation}
Here $\Delta x=\Delta y=L/192$.
The four Lagrange weights are
\begin{equation}
\begin{aligned}
 \ell_{-1}(s)&=-s(s-1)(s-2)/6,&\ell_0(s)&=(s+1)(s-1)(s-2)/2,\\
 \ell_1(s)&=-(s+1)s(s-2)/2,&\ell_2(s)&=(s+1)s(s-1)/6.
\end{aligned}
\end{equation}
For a fluid quantity $F$, the interpolated value is
\begin{equation}
 F(x_p,y_p,t)\approx\sum_{a=-1}^{2}\sum_{b=-1}^{2}
 \ell_a(s_x)\ell_b(s_y)F(x_{i+a},y_{j+b},t),
\end{equation}
with grid indices wrapped periodically.
Using the fluid fields at $t_{n+1}$, we solve the coupled backward-Euler equations
\begin{equation}
\begin{aligned}
 \bm x_p^{n+1}&=\bm x_p^n+\Delta t\bm v_p^{n+1},\qquad
 \bm v_p^{n+1}=\frac{\bm v_p^n+(\Delta t/\tau_p^{n+1})\bm u(\bm x_p^{n+1},t_{n+1})+\Delta t\bm g}
 {1+\Delta t/\tau_p^{n+1}},\\
 m_p^{n+1}&=m_p^n+\Delta t\dot m_p^{n+1},\qquad
 0=\dot H_p^{n+1}+L_v\dot m_p^{n+1}.
\end{aligned}
\label{eq:IIbe}
\end{equation}
Here $\dot H_p=2\pi r_p\rho_ac_{pa}\kappa_T\mathrm{Nu}_p(T_p^{\rm amb}-T_p)$.
We solve the backward-Euler mass equation by Newton iteration, limiting each correction so that the mass decreases by at most half its current value, while enforcing the heat balance.
Fixed-radius particles solve the first two equations with constant $\tau_p$.
The periodic coordinates and integer winding numbers are updated from the converged displacement.
Writing $y_p\in[0,L)$ and $k_p\in\mathbb Z$, the unwrapped height is $y_p^{\rm unwrapped}=y_p+Lk_p$.
Interpolation uses the periodic position; the background in Eq.~\eqref{eq:IIambient} uses the unwrapped height.

\subsubsection{Preferential-concentration statistics}
We reconstruct the concentration and particle moments using the quintic B-spline kernel in Eqs.~\eqref{eq:suppkernel} and \eqref{eq:suppkernelmoments}, with width $h=1.276\,\mathrm{mm}$ setting the observation scale.
For a uniform particle distribution, the square kernel support of side $6h$ contains approximately $N(6h/L)^2=384$ particles.
We evaluate these fields at the cell centers of a uniform $512^2$ Cartesian quadrature grid, whose spacing controls spatial integration accuracy at fixed $h$.
Each moment and its spatial derivatives are computed by summing particle values weighted by the kernel and its analytical derivatives.
Points outside all kernel supports contribute zero to the integrals.
The relative entropy and its contributions in Part I are evaluated on the quadrature grid and integrated by the midpoint rule,
\begin{equation}
 \int F(x,y)\,\mathrm dx\,\mathrm dy
 \simeq\Delta x^2\sum_{i,j}F((i+1/2)\Delta x,(j+1/2)\Delta x),\qquad \Delta x=L/512.
 \label{eq:IIquadrature}
\end{equation}
For the strain approximation in Eq.~\eqref{eq:averagingresolvedstrainrate}, fluid velocity gradients are computed by spectral differentiation and interpolated to the same quadrature points using four-point Lagrange interpolation.

We integrate each contribution rate by the trapezoidal rule to obtain its cumulative contribution to the relative-entropy change.
The settling-history contributions are obtained by applying the same rule twice in succession.
Entropy rates are evaluated every $0.5\,\mathrm{ms}$ time step and history sources every $2.5\,\mathrm{ms}$.

\subsection{Verification and validation}
\label{sec:IIverification}
\subsubsection{Validity of the small-inertia approximation}
We evaluate the timescale ratios and velocity remainder in Eq.~\eqref{eq:slipremainder} using the main calculation.
Over all droplets and saved times in $0\le t\le2\,\mathrm s$, the maximum of $\tau_p\|\nabla\bm u\|_F$ is $0.0506$, where $\|\cdot\|_F$ is the Frobenius norm.
Thus, the response time is at most $5.06\%$ of the local velocity-gradient timescale $1/\|\nabla\bm u\|_F$.
The ratio of the response time to its evolution timescale $\tau_p/|\dot\tau_p|$ is $|\dot\tau_p|$, whose maximum is $1.54\times10^{-3}$.
At $t=1$ and $2\,\mathrm s$, the maximum of $\tau_p|D_p\bm b|/|\bm b|$ is $1.51\times10^{-3}$ and $8.18\times10^{-4}$, respectively, confirming that the response time is also short compared with the variation timescale of $\bm b=\bm g-\bm a$ along droplet trajectories.
We quantify the velocity approximation error relative to the first-order correction $\tau_p\bm b$.
The maximum of $|\bm e_p|/|\tau_p\bm b|$ over all droplets is $2.02\%$ at $t=1\,\mathrm s$ and $2.24\%$ at $t=2\,\mathrm s$.
The cumulative contribution of the velocity remainder to the relative-entropy increase at $2\,\mathrm s$ is $\mathcal D_{\mathrm{rem}}=-1.19\times10^{-4}$, only $0.075\%$ of that increase in magnitude.
These results support the small-inertia approximation and its use in the entropy budget.

\begin{figure}[!b]
\centering\input{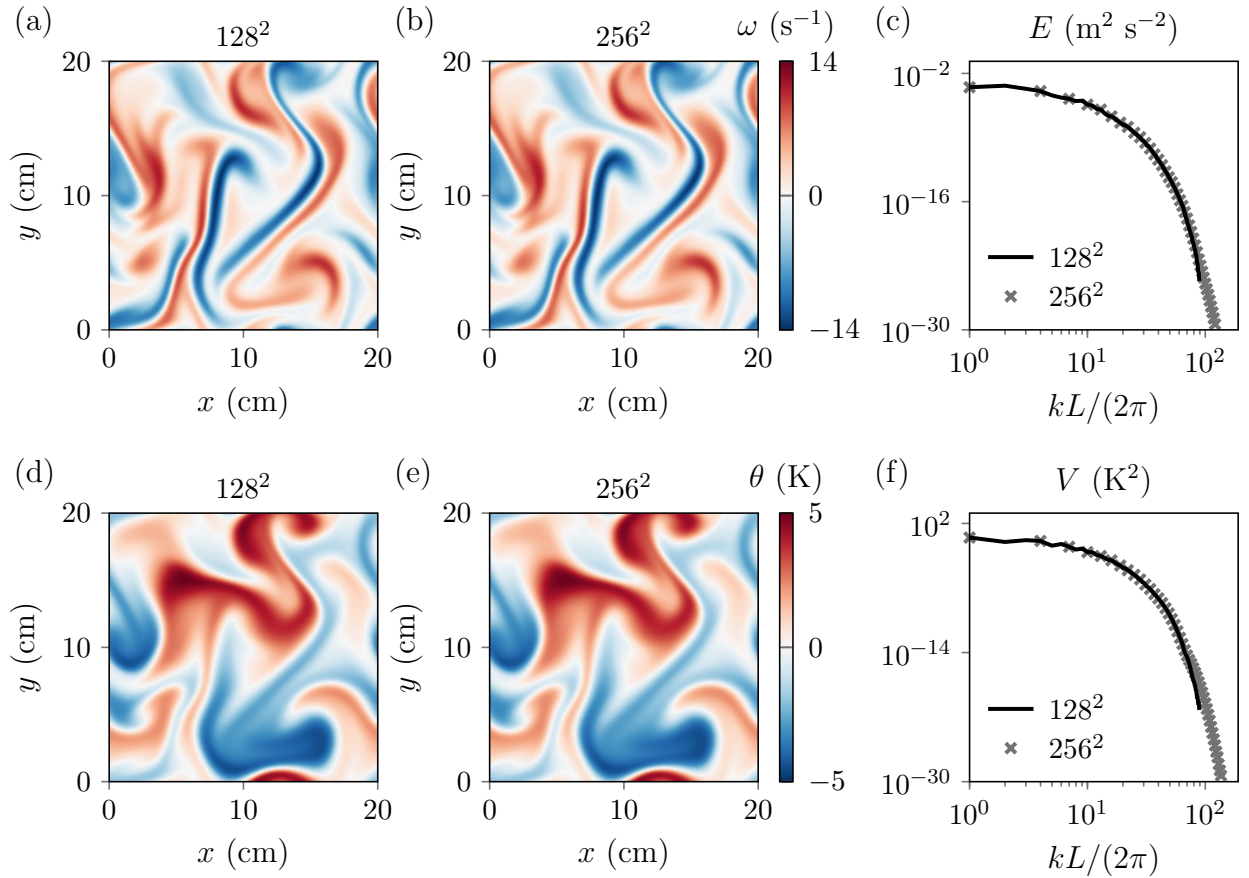}
\par\nointerlineskip
\setlength{\abovecaptionskip}{0pt}
\caption{(a,b) Vorticity and (d,e) temperature fluctuation at $t=2\,\mathrm s$ for $128^2$ and $256^2$ Fourier modes. (c) Kinetic-energy spectrum $E$ and (f) temperature-variance spectrum $V$.}
\label{fig:IIfluidresolution}
\end{figure}
\subsubsection{Fluid spatial resolution}
We compare simulations with $128^2$ and $256^2$ Fourier modes, initialized from the same fluid state at particle release and advanced independently for $2\,\mathrm s$.
The finer initial field is obtained by zero-padding the Fourier coefficients.
Figure~\ref{fig:IIfluidresolution} compares the vorticity and temperature fields, together with the kinetic-energy and temperature-variance spectra, at $t=2\,\mathrm s$.
The instantaneous fields and spectra agree very well between the two resolutions.

\subsubsection{Quadrature grid}
\begin{figure}[!t]
\centering\input{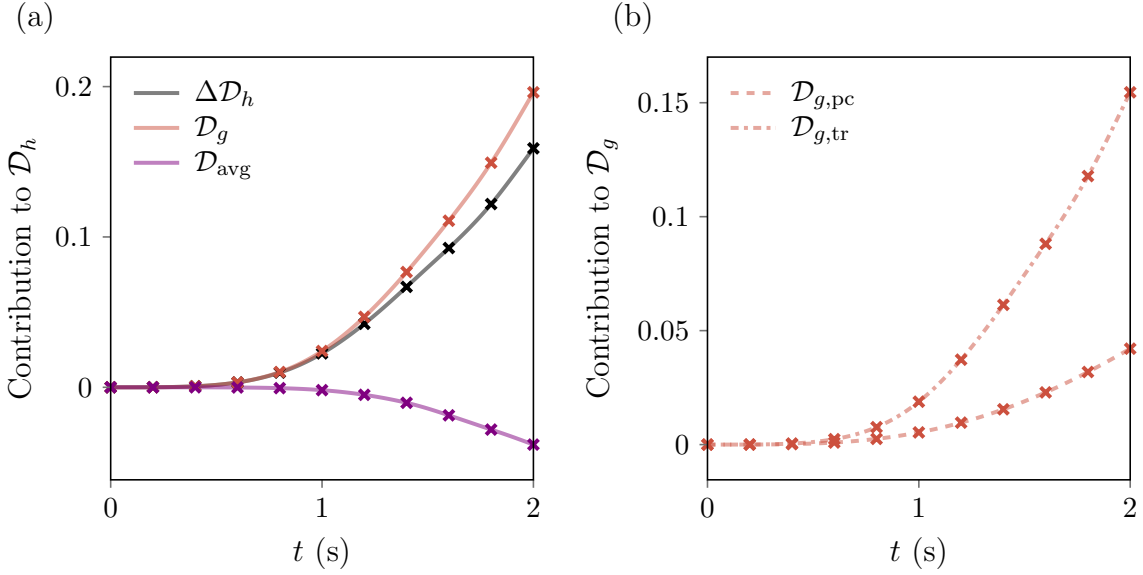}
\setlength{\abovecaptionskip}{0pt}
\caption{(a) Relative-entropy increase and its settling and spatial-averaging contributions. (b) Phase-change and transport contributions to the settling history. Lines and crosses, with $50\%$ and $100\%$ opacity, show $256^2$ and $512^2$ quadrature points, respectively.}
\label{fig:IIquadrature}
\end{figure}

Figure~\ref{fig:IIquadrature} compares $256^2$ and $512^2$ quadrature points for the same $512^2$ particle trajectories and kernel width.
The curves for the relative entropy and its physical and history contributions collapse throughout the observation interval.
The largest absolute difference among these five cumulative quantities is $1.59\times10^{-6}$.
The small differences confirm that the quadrature grid adequately resolves these integrals.
We therefore use the $256^2$ quadrature grid for the particle-number and time-step comparisons below.

\subsubsection{Particle number}
\begin{figure}[!htbp]
\centering\input{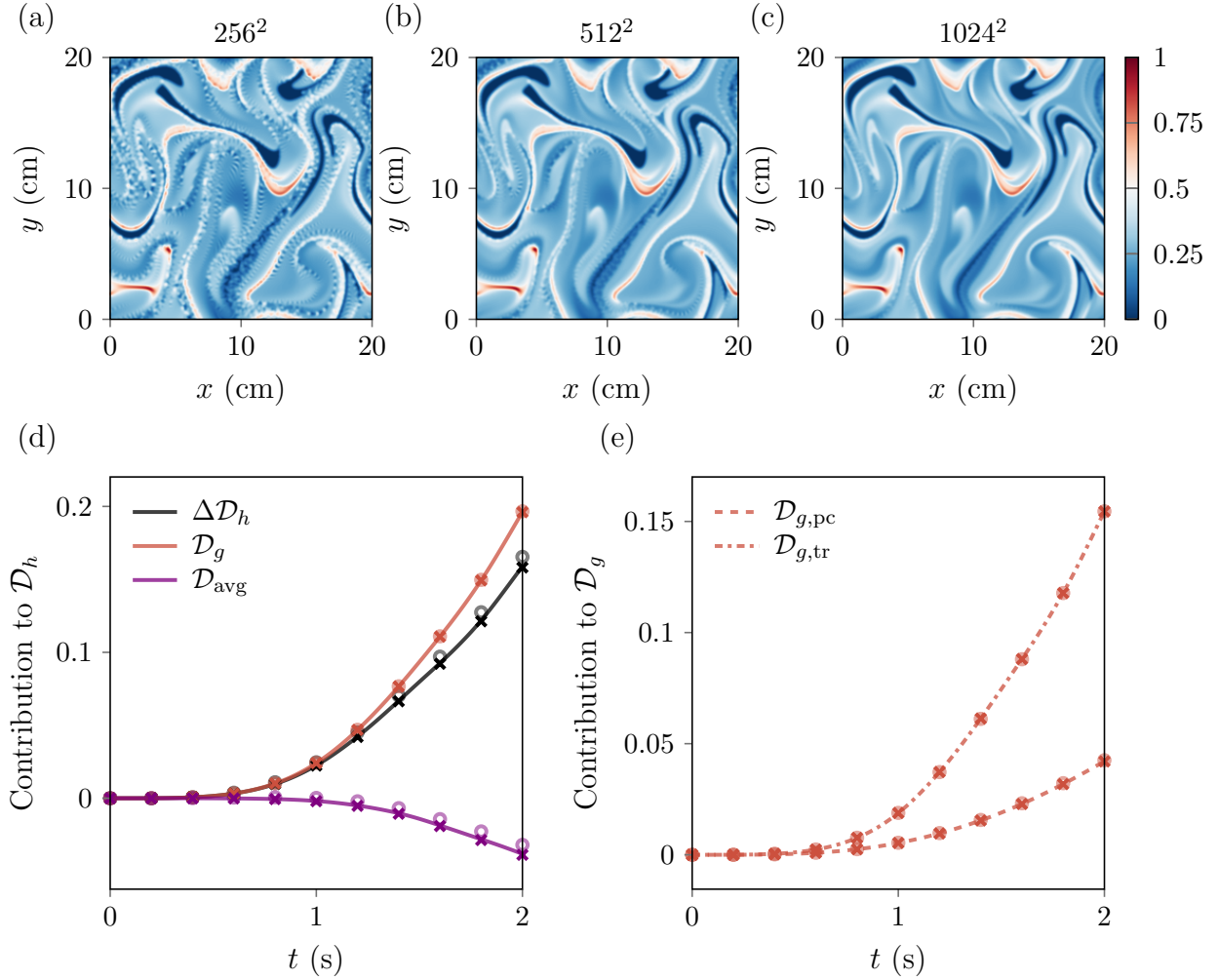}
\par\nointerlineskip
\setlength{\abovecaptionskip}{0pt}
\caption{(a--c) Phase-changing droplet concentration at $t=2\,\mathrm s$ for $256^2$, $512^2$, and $1024^2$ particles per population. Colors show $\log_{10}[1+c_h/(N/L^2)]$. (d) Relative-entropy increase and its settling and spatial-averaging contributions. (e) Phase-change and transport contributions to the settling history. Circles, lines, and crosses represent the three increasing particle counts, with opacities of $50\%$, $75\%$, and $100\%$, respectively.}
\label{fig:IIparticles}
\end{figure}

We next vary the number of Lagrangian particles per population from $256^2$ to $512^2$ and $1024^2$, keeping the flow, time step, kernel width, and $256^2$ quadrature grid fixed.
Figure~\ref{fig:IIparticles}(a--c) shows the same concentration structures, with more particles reducing the small-scale sampling variations.
The curves in panels (d,e) give the same physical conclusions over $0\le t\le2\,\mathrm s$.
Settling increases the relative entropy, spatial averaging reduces it, and transport contributes more than phase change to the settling history.
At $2\,\mathrm s$, the entropy-increase difference between phase-changing and fixed-radius particles varies by at most $0.09\%$ relative to the $512^2$-particle value.

\subsubsection{Time step}
\begin{figure}[!htbp]
\centering\input{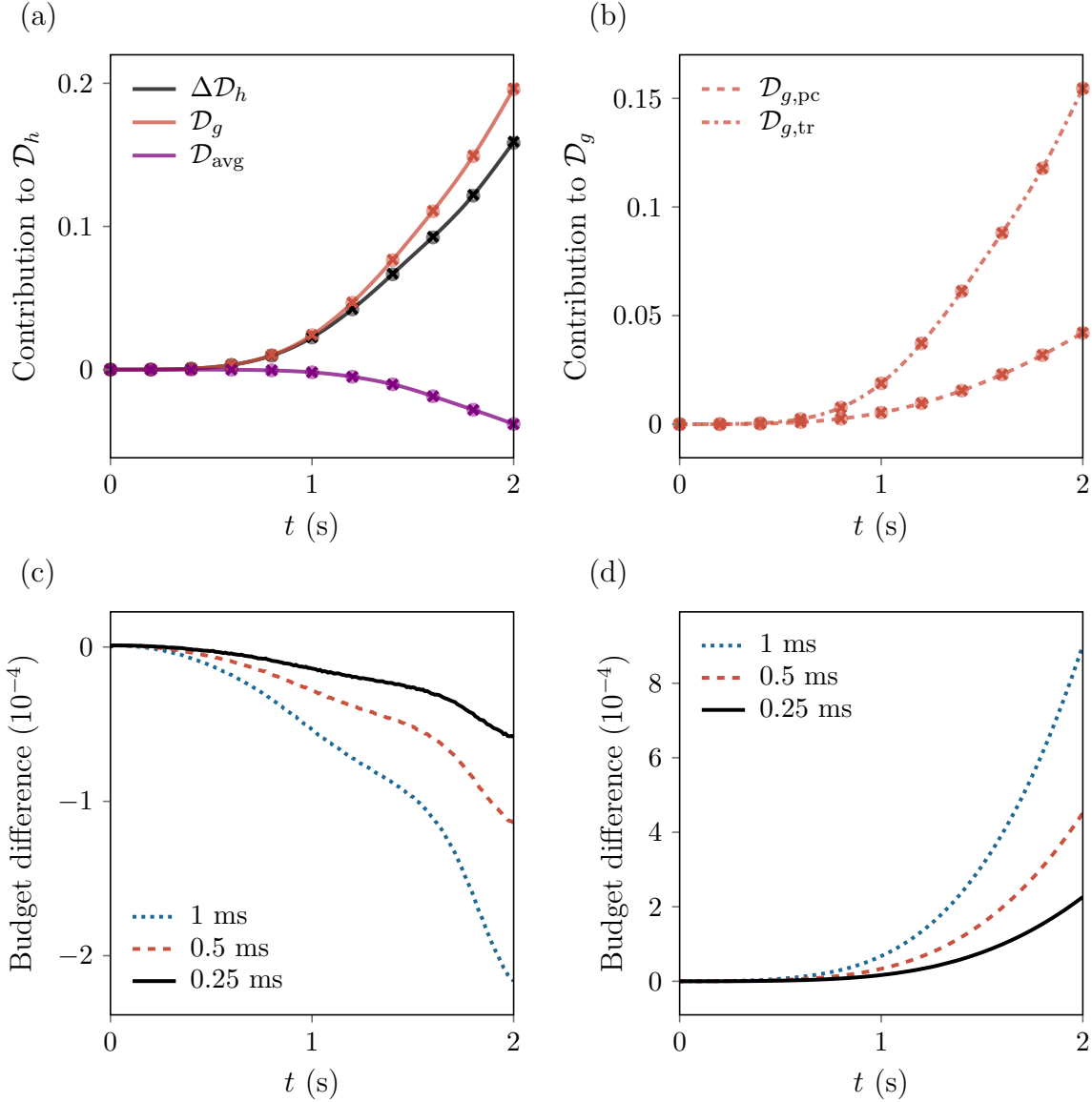}
\setlength{\abovecaptionskip}{0pt}
\caption{(a) Relative-entropy increase and its settling and spatial-averaging contributions. (b) Phase-change and transport contributions to the settling history. Circles, lines, and crosses show time steps of $1$, $0.5$, and $0.25\,\mathrm{ms}$, with opacities of $50\%$, $75\%$, and $100\%$, respectively. (c) Relative-entropy increase calculated from the concentration field minus the accumulated flux budget, including the discrete operator residual. (d) Settling-history budget minus the directly integrated settling contribution.}%
\label{fig:IIclosure}
\end{figure}

Figure~\ref{fig:IIclosure}(a,b) compares $\Delta t=1$, $0.5$, and $0.25\,\mathrm{ms}$ at fixed $512^2$ particles and $256^2$ quadrature points.
The curves approach one another as the step decreases, while the contribution ordering is unchanged.
Panels (c,d) show the differences between the directly evaluated quantities and their respective budgets; these decrease approximately in proportion to the time step, consistent with first-order particle time integration.
The entropy budget includes the discrete operator residual, defined as the entropy rate evaluated from $c_h[\bm u+\langle\tau\rangle_h(\bm g-\bm a)]$ minus $\dot{\mathcal D}_Q+\dot{\mathcal D}_g+\dot{\mathcal D}_a$.

\begin{figure}[!b]
\centering\input{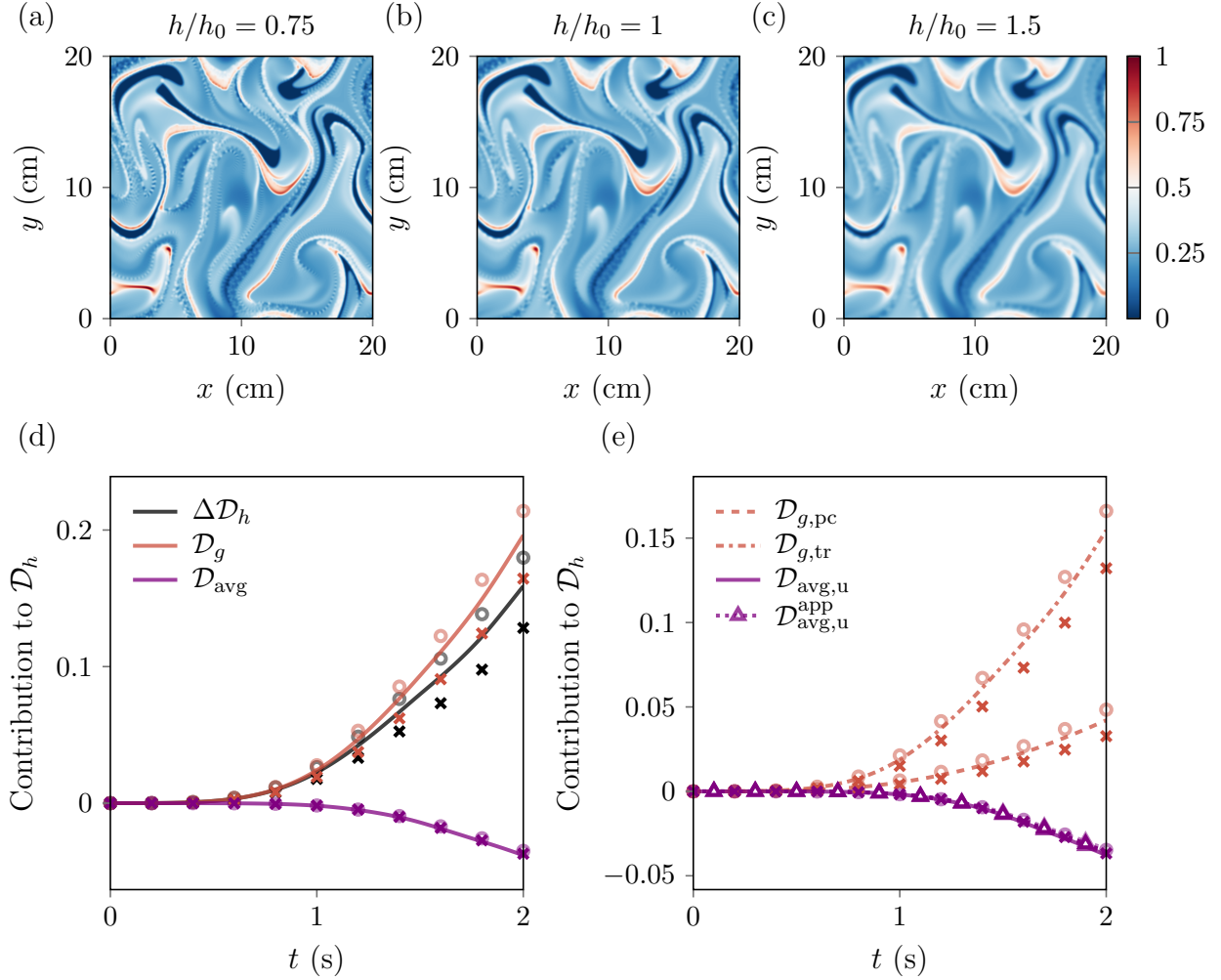}
\par\nointerlineskip
\setlength{\abovecaptionskip}{0pt}
\caption{(a--c) Phase-changing droplet concentration at $t=2\,\mathrm s$ for $h/h_0=0.75$, $1$, and $1.5$. Colors show $\log_{10}[1+c_h/(N/L^2)]$. (d) Relative-entropy increase and its settling and spatial-averaging contributions. (e) Phase-change and transport contributions to the settling history, together with the velocity-averaging contribution and its strain approximation. $\mathcal D_{\mathrm{avg},u}^{\mathrm{app}}$ is shown by dotted lines with triangles; other quantities use circles, unmarked lines, and crosses for the three increasing kernel widths. Opacities of $50\%$, $75\%$, and $100\%$ distinguish these widths throughout (d,e).}
\label{fig:IIwidth}
\end{figure}
\subsubsection{Kernel width}
The kernel width sets the observation scale and is varied independently of numerical resolution.
We denote the kernel width used in the main calculation by $h_0=1.276\,\mathrm{mm}$.
Figure~\ref{fig:IIwidth} compares $h/h_0=0.75$, $1$, and $1.5$ on the same trajectories.
Increasing the width smooths the concentration structures and slightly reduces the entropy increase, while settling remains the dominant positive contribution, spatial averaging offsets part of it, and transport exceeds the phase-change history contribution.
The direct velocity-averaging contribution and its strain approximation also agree throughout the interval [Fig.~\ref{fig:IIwidth}(e)].

\subsubsection{Release time}
\begin{figure}[!htbp]
\centering\input{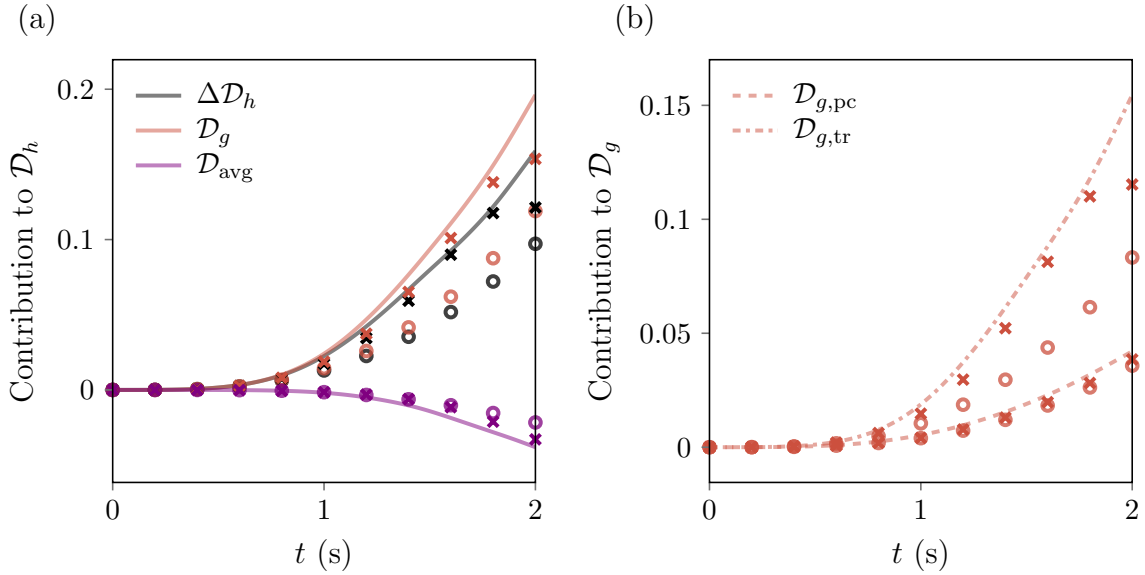}
\setlength{\abovecaptionskip}{0pt}
\caption{(a) Relative-entropy increase and its settling and spatial-averaging contributions. (b) Phase-change and transport contributions to the settling history. Lines, circles, and crosses represent the three increasing release times, with opacities of \(50\%\), \(75\%\), and \(100\%\), respectively.}
\label{fig:IIrelease}
\end{figure}

Figure~\ref{fig:IIrelease} compares release at $20t_f$, $30t_f$, and $40t_f$ after flow initialization, with time measured from each release.
Droplets released at different stages of flow evolution experience different thermodynamic histories, yet settling remains the dominant positive contribution and transport contributes more to the settling history than phase change.

\begin{figure}[!p]
\centering\begin{tikzpicture}
\definecolor{densityZero}{RGB}{5,48,97}\definecolor{strainMauve}{HTML}{B34A98}\begin{axis}[paper axis,name=p00,
at={(0.0\linewidth,.73\linewidth)},anchor=south west,
width=0.3\linewidth,height=0.3\linewidth,
axis background/.style={fill=densityZero},xmin=0,xmax=20,ymin=0,ymax=20.0,xtick={0,10,20},xlabel={$x$ (cm)},ylabel={$y$ (cm)},ytick={0,10,20},]
\addplot graphics[xmin=0,xmax=20,ymin=-2.710505431213761e-18,ymax=20.0] {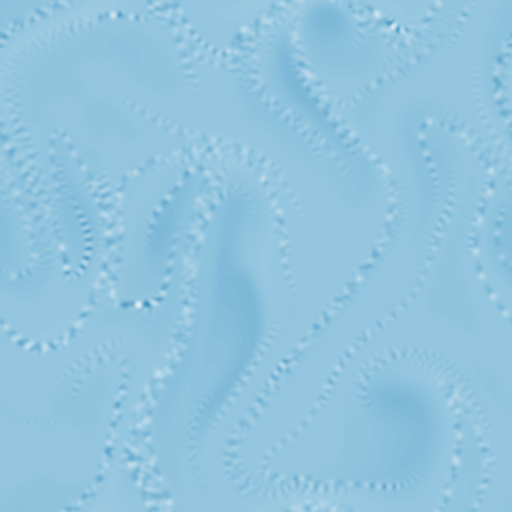};
\node[font=\small,anchor=south] at (rel axis cs:.5,1.02) {Fixed-radius};
\end{axis}
\node[anchor=south west,font=\normalsize] at ([xshift=-0.085\linewidth,yshift=5pt]p00.north west) {(a)};
\begin{axis}[paper axis,name=p10,
at={(0.0\linewidth,0pt)},anchor=south west,
width=0.3\linewidth,height=0.6\linewidth,
axis background/.style={fill=densityZero},xmin=0,xmax=20,ymin=-15.0,ymax=25.0,xtick={0,10,20},xlabel={$x$ (cm)},ylabel={$y$ (cm)},ytick={-10,0,10,20},]
\addplot graphics[xmin=0,xmax=20,ymin=-13.945312500000002,ymax=24.570312500000004] {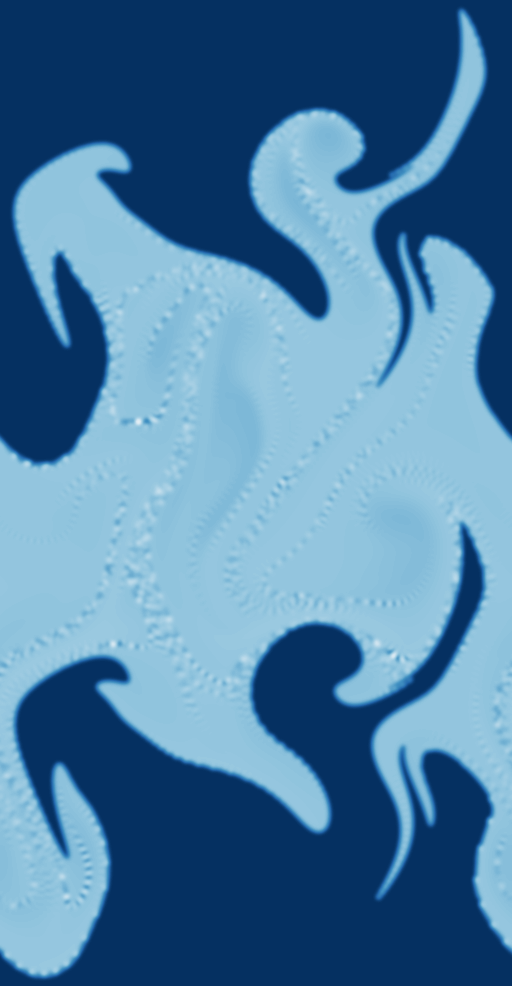};

\end{axis}
\node[anchor=south west,font=\normalsize] at ([xshift=-0.085\linewidth,yshift=5pt]p10.north west) {(c)};
\begin{axis}[paper axis,name=p01,
at={(0.48\linewidth,.73\linewidth)},anchor=south west,
width=0.3\linewidth,height=0.3\linewidth,
axis background/.style={fill=densityZero},xmin=0,xmax=20,ymin=0,ymax=20.0,xtick={0,10,20},xlabel={$x$ (cm)},ylabel={$y$ (cm)},ytick={0,10,20},colormap={density}{rgb255(0cm)=(5,48,97);rgb255(1cm)=(22,81,144);rgb255(2cm)=(41,113,177);rgb255(3cm)=(63,141,192);rgb255(4cm)=(107,172,208);rgb255(5cm)=(154,201,224);rgb255(6cm)=(194,221,235);rgb255(7cm)=(223,236,242);rgb255(8cm)=(247,246,246);rgb255(9cm)=(250,228,215);rgb255(10cm)=(250,204,180);rgb255(11cm)=(244,170,136);rgb255(12cm)=(228,128,101);rgb255(13cm)=(208,84,71);rgb255(14cm)=(185,39,50);rgb255(15cm)=(147,14,38);rgb255(16cm)=(103,0,31)},point meta min=0,point meta max=1,colorbar,colorbar style={at={(1.08,0)},anchor=south west,width=5pt,height=.30\linewidth,ytick={0,.25,.5,.75,1},axis line style={line width=.6pt},yticklabel style={xshift=2pt}},]
\addplot graphics[xmin=0,xmax=20,ymin=-2.710505431213761e-18,ymax=20.0] {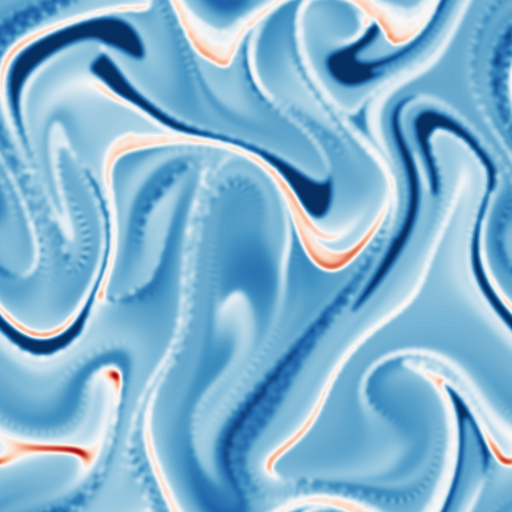};
\node[font=\small,anchor=south] at (rel axis cs:.5,1.02) {Phase-changing};
\end{axis}
\node[anchor=south west,font=\normalsize] at ([xshift=-0.085\linewidth,yshift=5pt]p01.north west) {(b)};
\begin{axis}[paper axis,name=p11,
at={(0.48\linewidth,0pt)},anchor=south west,
width=0.3\linewidth,height=0.6\linewidth,
axis background/.style={fill=densityZero},xmin=0,xmax=20,ymin=-15.0,ymax=25.0,xtick={0,10,20},xlabel={$x$ (cm)},ylabel={$y$ (cm)},ytick={-10,0,10,20},]
\addplot graphics[xmin=0,xmax=20,ymin=-13.554687500000002,ymax=24.062499999999996] {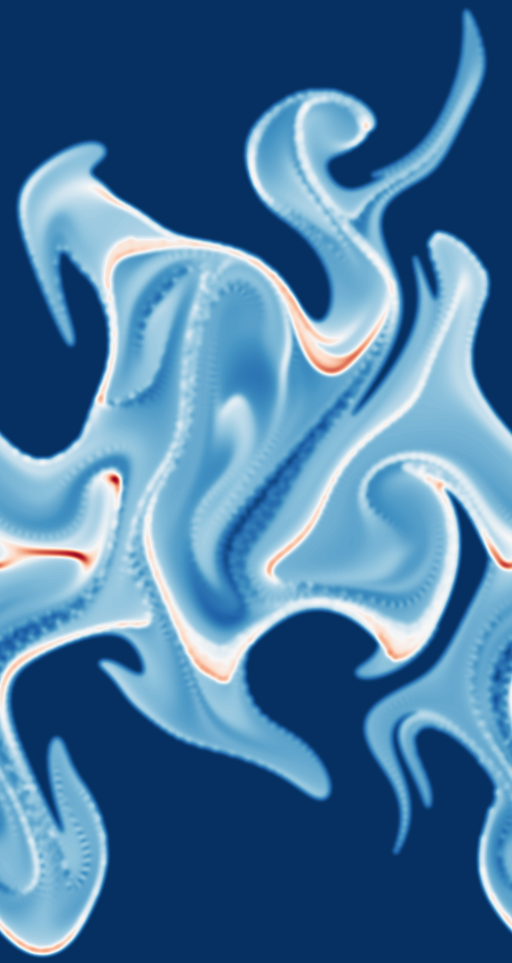};

\end{axis}
\node[anchor=south west,font=\normalsize] at ([xshift=-0.085\linewidth,yshift=5pt]p11.north west) {(d)};
\draw[strainMauve,line width=1.5pt] ([yshift=0.225\linewidth]p10.south west) rectangle ([yshift=0.525\linewidth]p10.south east);\draw[strainMauve,line width=1.5pt] ([yshift=0.225\linewidth]p11.south west) rectangle ([yshift=0.525\linewidth]p11.south east);\end{tikzpicture}
\setlength{\abovecaptionskip}{0pt}
\caption{(a,b) Concentration after periodic projection and (c,d) concentration at the unwrapped droplet heights at $t=2\,\mathrm s$. The left and right columns show fixed-radius particles and phase-changing droplets, respectively. Colors show $\log_{10}[1+c_h/(N/L^2)]$. Both descriptions use the same $512^2$ trajectories per population. The rectangles in (c,d) mark the initial square $[0,L]\times[0,L]$.}
\label{fig:IIcoordinates}
\end{figure}
\subsubsection{Unwrapped height and periodic projection}
\begin{figure}[!b]
\centering\input{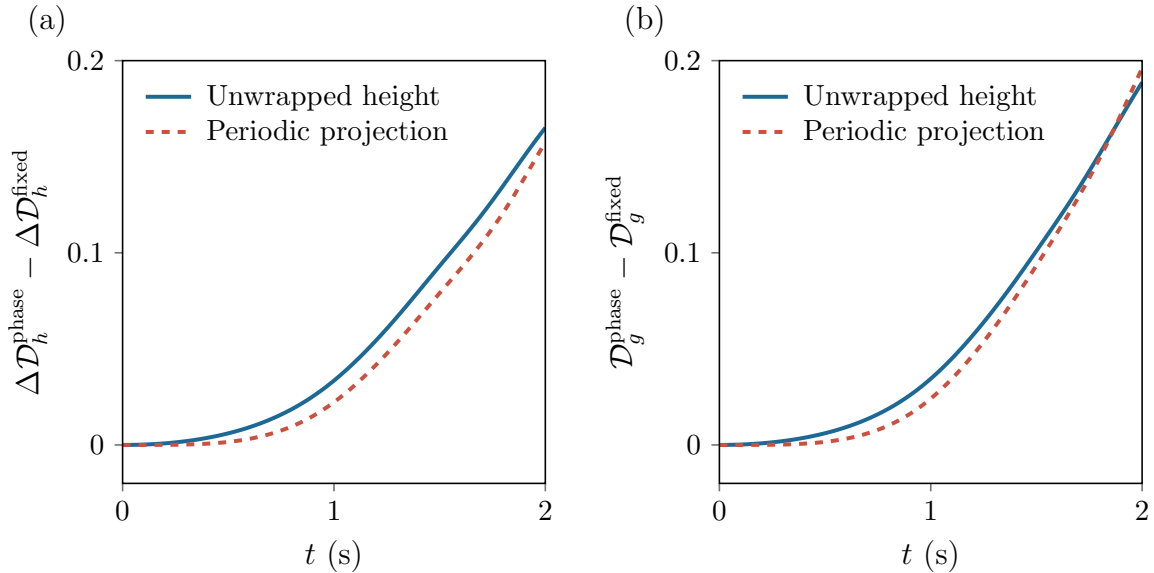}
\setlength{\abovecaptionskip}{0pt}
\caption{(a) Difference in relative-entropy increase, $\Delta\mathcal D_h^{\mathrm{phase}}-\Delta\mathcal D_h^{\mathrm{fixed}}$, and (b) difference in cumulative settling contributions, $\mathcal D_g^{\mathrm{phase}}-\mathcal D_g^{\mathrm{fixed}}$.}
\label{fig:IIcoordinatecurves}
\end{figure}

We compare concentration at the particles' unwrapped heights with concentration after vertical periodic projection, using the same trajectories.
Figure~\ref{fig:IIcoordinates} shows both populations at $2\,\mathrm s$.
Even without periodic projection, phase-changing droplets exhibit pronounced local preferential concentration within the droplet layer, whereas fixed-radius particles remain more uniformly distributed. The difference between the two populations is therefore evident in both coordinate descriptions.
To evaluate concentration at the unwrapped droplet heights, we extend the statistical integration region vertically while retaining the original periodic flow.
The kernel is periodic in $x$ and uses the continuous distance in $y$.
Online integration uses the fixed region $0\le x<L$, $-8L\le y\le5L$, with $256\times3328$ quadrature points and $\Delta x=\Delta y=L/256$.
For $L=0.20\,\mathrm m$, the vertical limits are $-1.6$ and $1.0\,\mathrm m$.
We calculate
\begin{equation}
 \mathcal D_h^{\mathrm{unwrapped}}(t)=\frac1N\int_{-8L}^{5L}\!\int_0^L c_h\ln\frac{13L^2c_h}{N}\,\mathrm dx\,\mathrm dy
 \simeq\frac{\Delta x\Delta y}{N}\sum_{i,j}c_{h,ij}\ln\frac{13L^2c_{h,ij}}{N}.
 \label{eq:IIactualheight}
\end{equation}
The reference concentration is $N/(13L^2)$, uniform over the integration region, and we use $0\ln 0=0$.
The integration region contains all particle kernel supports throughout $0\le t\le2\,\mathrm s$.
Figure~\ref{fig:IIcoordinatecurves} compares $\Delta\mathcal D_h^{\mathrm{phase}}(t)-\Delta\mathcal D_h^{\mathrm{fixed}}(t)$ and $\mathcal D_g^{\mathrm{phase}}(t)-\mathcal D_g^{\mathrm{fixed}}(t)$, where $\Delta\mathcal D_h(t)=\mathcal D_h(t)-\mathcal D_h(0)$ and the superscripts $\mathrm{phase}$ and $\mathrm{fixed}$ denote phase-changing droplets and fixed-radius particles.
Both differences are evaluated using either unwrapped heights or periodic projection; for the former, $\mathcal D_h$ is replaced by $\mathcal D_h^{\mathrm{unwrapped}}$.
At $2\,\mathrm s$, the entropy-increase differences are $0.16509$ and $0.15768$, respectively; the settling differences are $0.18869$ and $0.19620$.
Both coordinate descriptions therefore retain the enhanced concentration nonuniformity and dominant settling contribution over the main observation interval.

\end{document}